\documentclass[aps,prd,nofootinbib,onecolumn,notitlepage,10pt]{revtex4-1}
\usepackage[utf8]{inputenc}
\usepackage{amsmath}
\usepackage{amsfonts}
\usepackage{amssymb}
\usepackage{color}
\usepackage[unicode=true,pdfusetitle,
 bookmarks=true,bookmarksnumbered=false,bookmarksopen=false,
 breaklinks=false,backref=false,colorlinks=false]
 {hyperref}

\usepackage{graphicx}
\usepackage{dcolumn}
\usepackage{bm}
\usepackage{hyperref}

\usepackage{bbm}
\usepackage{xcolor}
\usepackage[normalem]{ulem}

\newcommand{\LF}{\left(}
\newcommand{\RF}{\right)}
\newcommand{\LT}{\left[}
\newcommand{\RT}{\right]}
\newcommand{\pd}{\partial}
\newcommand{\Pc}{\mathcal{P}}
\newcommand{\Tc}{\mathcal{T}}
\newcommand{\Hc}{\mathcal{H}}
\newcommand{\Fc}{\mathcal{F}}
\newcommand{\Oc}{\mathcal{O}}
\newcommand{\Lc}{\mathcal{L}}
\newcommand{\Uc}{\mathcal{U}}
\newcommand{\Vc}{\mathcal{V}}
\newcommand{\Cc}{\mathcal{C}}

\begin{document}


\title{Revisiting quantum field theory in Rindler spacetime with superselection rules}

\author{K. Sravan Kumar}
\email{sravan.kumar@port.ac.uk}
 \affiliation{Institute of Cosmology \& Gravitation,
	University of Portsmouth,
	Dennis Sciama Building, Burnaby Road,
	Portsmouth, PO1 3FX, United Kingdom}
\author{Jo\~ao Marto}%
 \email{jmarto@ubi.pt}
\affiliation{%
Departamento de F\'isica, Centro de Matem\'atica e Aplicações (CMA-UBI), Universidade da Beira Interior, Rua Marquês D'Ávila e Bolama, 6201-001 Covilhã, Portugal}%

\date{\today}

\begin{abstract}
Quantum field theory (QFT) in Rindler spacetime is a gateway to understanding unitarity and information loss paradoxes in curved spacetime. Rindler coordinates map Minkowski spacetime onto regions with horizons, effectively dividing accelerated observers into causally disconnected sectors. Employing standard quantum field theory techniques and Bogoliubov transformations between Minkowski and Rindler coordinates yields entanglement between states across these causally separated regions of spacetime. This results in a breakdown of unitarity, implying that information regarding the entangled partner may be irretrievably lost beyond the Rindler horizon. As a consequence, one has a situation of pure states evolving into mixed states. In this paper, we introduce a novel framework for comprehending this phenomenon using a recently proposed formulation of direct-sum quantum field theory (DQFT), which is grounded in superselection rules formulated by the parity and time reversal ($\Pc\Tc$) symmetry of Minkowski spacetime. In the context of DQFT applied to Rindler spacetime, we demonstrate that each Rindler observer can, in principle, access pure states within the horizon, thereby restoring unitarity. However, our analysis also reveals the emergence of a thermal spectrum of Unruh radiation. This prompts a reevaluation of entanglement in Rindler spacetime, where we propose a novel perspective on how Rindler observers may reconstruct complementary information beyond the horizon. Furthermore, we revisit the implications of the Reeh-Schlieder theorem within the framework of DQFT. Lastly, we underscore how our findings contribute to ongoing efforts aimed at elucidating the role of unitarity in quantum field theory within the context of de Sitter and black hole spacetimes.
\end{abstract}

\maketitle

\section{Introduction}

Hawking's seminal paper on black hole radiation \cite{Hawking:1974sw} posed very concerning questions on our understanding of gravity and quantum mechanics. The question is about the loss of unitarity, which is an observer losing a part of the quantum information behind the horizon, and it manifested in general into the well-known information loss paradox. The extrapolation of the information loss paradox is Hawking radiation being independent of what has formed the black hole. There are a few setbacks in this extrapolation, such that, after all these years, we still do not have a consistent understanding of quantum field theory (QFT) in curved spacetime, especially when spacetime is dynamical in nature. In particular, since we do not yet know the consistent framework of quantum fields in a dynamical collapse, the microscopic understanding of black hole formation is an open question. Going to the details of Hawking's calculation, one can decipher that the quantization is performed in Schwarzchild spacetime, which is static in {nature} 
(the detailed assumptions in Hawking's derivation can be found in \cite{Kumar:2023hbj}).  Soon after Hawking's paper, Unruh's remarkable calculation of QFT in Rindler spacetime \cite{Unruh1976@article,Higuchi:2017gcd,Crispino:2007eb,Mukhanov:2007zz} reaffirms the unitarity loss due to the ignorance of the entangled state part beyond the Rindler horizon. Numerous attempts have been made over the decades to resolve the problem in the context of several frameworks of quantum gravity \cite{Raju:2020smc,Almheiri_2020}, which include Hawking's last works on this \mbox{subject \cite{Hawking:2016msc}.} The seminal papers of Norma G. Sanch\'ez, \mbox{B. Whiting} and \mbox{Gerard 't Hooft} have taken forward the earliest idea of Schr\"{o}dinger's antipodal identification to resolve the unitarity and information loss issues \cite{SANCHEZ1987605,SANCHEZ19871111,Schrodinger1956,tHooft:2016rrl,tHooft:1993dmi,tHooft:2018waj}. 
 Despite all these attempts, a fundamental question one can ask here is if one has to reanalyze the historical developments and carefully scrutinize each step of the calculation that was performed and find if the conceptual conundrums are mere artifacts of the way we perceive quantum theory when there are spacetime horizons. This paper is such an attempt. In our earlier attempts we have developed the subject of QFT in curved spacetime towards achieving unitarity using superselection rules \cite{nlab:superselection_theory,Kumar:2023ctp,Kumar:2023hbj}, that constructs a global Hilbert space as direct-sum of Hiblert spaces which describe parity conjugate regions of physical space. We have shown in these papers how one can achieve unitarity (i.e., the evolution of pure states into pure states) if one starts understanding (quantum) spacetime with parity ($\Pc$) and time \mbox{reversal ($\Tc$)} operations. In this paper, we address the QFT in Rindler spacetime in the context of direct-sum QFT (QDFT) developed in \cite{Kumar:2023ctp}. We provide further support for DQFT's framework in uncovering the fundamental conundrums on our understanding of quantum fields in conjunction with spacetime horizons, which is very crucial for future endeavours for quantum gravity. 

The paper is organized as follows. In Section~\ref{sec:DQM}, we discuss the $\Pc\Tc$ symmetric formulation of quantum mechanics (QM) \cite{Kumar:2023ctp,Gaztanaga:2024vtr} that splits the single-particle Hilbert space into superselection sectors by $\Pc\Tc$ operation. In Section ~\ref{sec:DQFT} we discuss the \mbox{direct-sum QFT \cite{Kumar:2023ctp}} that again builds the QFT with $\Pc\Tc$ where vacuum and Fock space are constructed by superselection rules. In Section~\ref{sec:DQFTRindler}, we discuss the 1 + 1 dimensional Rindler spacetime that emerges by coordinate transformations on the Minkowski spacetime coordinates. These coordinate transformations confine the Minkowski spacetime into four regions perceived by constantly accelerating observers. We perform the calculations of how the Minkowski vacuum of DQFT would lead to (thermal) particle spectrum to the Rindler observers. Still, in our case, we show that unitarity can be maintained. In Section~\ref{sec:RStheoremDQFT}, we discuss the status of entanglement and the ramifications for the Reeh-Shlieder theorem in DQFT. In Section~\ref{sec:EqntanglePure}, we present how DQFT in Rindler spacetime leads to thermal spectrum for all Past, Future, Left, and Right Rindler regions, but still, one can preserve the evolution of pure states into pure states (unitarity) for every observer. In Section~\ref{sec:DifferencesQFTs}, we discuss the differences between the Unruh radiation in standard QFT and DQFT. We also qualitatively elucidate the merits of DQFT as an essential mechanism to get a consistent quantum theory and achieve observer complementarity. In Section~\ref{sec:unitaritycurvedST}, we comment on the connection between DQFT in Rindler spacetime to DQFT in curved spacetimes such as de Sitter and Schwarzchild that were explored in earlier works \cite{Kumar:2023ctp,Kumar:2023hbj}. In Section~\ref{sec:conc} we summarize the results with conceptual discussion. In Appendix~\ref{AppPT}, we provide details $\Pc\Tc$ symmetry in classical harmonic oscillator and contrast with notions considered in standard QM. We highlight the crucial aspects of building direct-sum QM. In Appendix~\ref{appbogo} we provide details of Bogoliubov transformations and the coventions we followed from \cite{Mukhanov:2007zz}.

Throughout the paper, we follow the natural units $\hbar=c=1$ and mostly positive metric $\LF -, +,\,+,\,+ \RF$ signature. 

\section{Direct-sum quantum mechanics}

\label{sec:DQM}

Direct-sum QM emerges as an alternative framework that offers a fresh perspective on quantum theory, especially in situations involving superselection rules. Unlike the standard formulation of quantum mechanics, which assumes a single global Hilbert space for describing the entire system, direct-sum QM decomposes the Hilbert space into a direct-sum of sectorial Hilbert spaces associated with different superselection sectors.
Superselection sectors arise in quantum mechanics when certain observables cannot be simultaneously measured with arbitrary precision.  From the perspective of an observer, superselection sectors represent distinct non-superimposable physical states that cannot be transformed into each other through local operations. Discrete spacetime transformations such as Parity ($\Pc$) and Time reversal ($\Tc$) fall into global operations. In the context of quantum mechanics, the operation of time reversal is typically represented by an anti-unitary operator acting on the quantum states of the system \cite{Simon_2008,Roberts:2022xcj}. This operator applies a complex conjugation operation to the wave function, effectively reversing the sign of the time parameter. 
It is important to note that \cite{Srednicki:2007qs,Coleman:2018mew}:

\begin{quote}
  {\it ``In QM, time is treated as a parameter where space has a status of operator. Even if we go to QFT, time remains as a parameter. QFT combines QM and special relativity by imposing the causality condition $\Big[\hat{\Oc}\LF x \RF,\, \hat{\Oc}\LF y \RF\Big]= 0$ which is the commutativity of operators for space-like distances $\LF x-y \RF^2>0$. ’’}
\end{quote}



In direct-sum QM, a single Hilbert space is divided into superselection sectors defined by the combined operation of $\Pc\Tc$ \cite{Kumar:2023ctp,Kumar:2023hbj,Gaztanaga:2024vtr}. Here, we review direct-sum QM, where a positive energy state is defined without reference to the arrow of time. 
{It is worthy of mentioning that the model presented in this paper, which is based on $\Pc\Tc$ symmetry and superselection sectors, bears some conceptual similarities to the ``Two-State Vector Formalism’’ (TSVF) of quantum mechanics reviewed in \cite{Aharonov2002}. While the ($\Pc\Tc$) symmetry and superselection sectors model and the Two-State Vector Formalism (TSVF) are distinct in their specific formulations and motivations, they share a conceptual affinity in terms of introducing additional structure and symmetries to the Hilbert space. The TSVF's emphasis on time symmetry and dual-state description offers a rich perspective that might provide useful analogies or methods for further developing and understanding our proposed $\Pc\Tc$ symmetry-based model. Exploring these connections could lead to deeper insights into the implications and applications of superselection sectors in quantum field theory.}
In usual QM (See Appendix~\ref{AppPT}), a positive energy state is defined by 
\begin{equation}
    \vert \Psi\rangle_t = e^{-iEt}\vert \Psi\rangle_0
\end{equation}
{with} 
 presumption on the arrow of time $t: -\infty \to \infty$. Another equivalent way of defining a positive energy state is 
\begin{equation}
    \vert \Psi\rangle_t = e^{iEt}\vert \Psi\rangle_0
\end{equation}
with presumption on the arrow of time $t: \infty \to -\infty$. Without referring to the arrow time, we can definite the positive energy state as 
\begin{equation}
    \vert \Psi\rangle = \frac{1}{\sqrt{2}} \LF \ \vert\Psi_+\rangle \oplus \vert \Psi_-\rangle \RF = \frac{1}{\sqrt{2}}\begin{pmatrix}
        \vert \Psi_+\rangle \\ 
        \vert \Psi_-\rangle 
    \end{pmatrix}
    \label{dss}
\end{equation}
where $\oplus$ is direct-sum operation and it makes the states $\vert\Psi_{\pm}\rangle $ orthogonal by \mbox{construction \cite{Conway}.} Note that it is different from the usual {summation 
(in the context of superposition of states, we use usual summation for example $\vert \psi\rangle = \vert\psi_1\rangle+\vert \psi_2\rangle$, whereas the direct-sum of two states increases the dimensionality of the state vector)}.
Here, $\vert \Psi_+\rangle$ is a positive energy state at a position (say $\textbf{x}$) (with an arrow of time $t:  -\infty \to \infty$) while $\vert \Psi_-\rangle$ being a positive energy state at a position $-\textbf{x}$ (with an arrow of time $t:  \infty \to -\infty$). This structurally implies the state \eqref{dss} is an invariant state under $\Pc\Tc$ symmetry, and it is governed by the direct-sum Schr\"{o}dinger equation 
\begin{equation}
    i\frac{\partial\vert \Psi\rangle}{\partial t_p} = \begin{pmatrix}
			\hat{{H}}_+ & 0 \\ 
			0 & -\hat{{H}}_-
		\end{pmatrix} \vert \Psi\rangle
  \label{disumSch}
\end{equation}
where $\hat{H}_\pm\LF \hat{x}_\pm,\,\hat{p}_\pm \RF$ give the full Hamiltonian of the system $\hat{H}=\hat{H}_+\oplus \hat{H}_-$, which are functions of position and momenta operators of the target space defined by
\begin{equation}
\begin{aligned}
\hat{p}_+ & = -i\frac{d}{dx_+},\quad x_+=x\gtrsim 0 \\ 
\hat{p}_- & = i\frac{d}{dx_-},\quad x_-=x\lesssim 0\,. 
\end{aligned}
\end{equation}
where the notation $\gtrsim$ and $\lesssim$ refer to the eigenvalues of $\hat x_\pm$ to be $x_+\in (0,\infty]$ and $x_-\in [-\infty, 0)$. 
{The} 
canonical non-zero commutation relations  
\begin{equation}
    [\hat{x}_+,\,\hat{p}_+] = i,\quad [\hat{x}_-,\,\hat{p}_-]=-i\,,
\end{equation}
{and the remaining relations}
\begin{equation}
\LT \hat x_+,\, \hat x_- \RT = \LT \hat p_+,\, \hat p_- \RT = \LT \hat x_+,\, \hat p_- \RT = \LT \hat p_+,\, \hat x_- \RT =0 \, ,
\end{equation}
define the Hilbert space 
\begin{equation}
    \Hc= \Hc_+ \oplus \Hc_-
    \label{SSSHilb}
\end{equation}
where the component states $\vert\Psi_\pm\rangle $ of $\vert \Psi\rangle$ \eqref{dss} evolve in $\Hc_\pm$ at the parity conjugate points in physical space. {The parity conjugate regions of physical space now define the Hilbert space with the superselection sectors $\Hc_{\pm}$ where positive energy states (components of a single state) evolve with opposite arrows of time  (See \cite{Kumar:2023ctp} for further details). It is important to note that the superselection rule here is based on $\Pc\Tc$ symmetry, which is structurally different and aesthetically similar to superselection rules applied in algebraic quantum field theory historically. 
The operation direct-sum $\oplus$ forbids any super-position of states among the superselection sectors defined by $\Pc\Tc$. 
The wave function is now defined by }
\begin{equation}
    \langle x\vert \Psi\rangle = \frac{1}{2} \begin{pmatrix}
       \langle x_+\vert \quad \langle x_- \vert 
    \end{pmatrix}\begin{pmatrix}
        \vert \Psi_+\rangle \\
        \vert\Psi_-\rangle  
    \end{pmatrix} = \frac{1}{2}  \Psi(x_+) e^{-iEt} + \frac{1}{2}  \Psi(x_-)e^{iEt}\,. 
    \label{wavefunc}
\end{equation}
{The} 
square integrabilities and probabilities in the total Hilbert space ($\Hc$) as well as the individual Hilbert spaces $\LF \Hc_+,\,\Hc_- \RF$ are sum up to unity as reflected by the \mbox{following equation}
\begin{equation}
  \int_{-\infty}^\infty dx  \langle \Psi\vert \Psi \rangle = \frac{1}{2} \int^{\infty}_0 dx_+  \langle \Psi_+\vert \Psi_+ \rangle + \frac{1}{2} \int^{0}_{-\infty} dx_-  \langle \Psi_-\vert \Psi_- \rangle  =1\,. 
\end{equation}
\subsection{Quantum Harmonic Oscillator}
\begin{figure}
\includegraphics[width=0.6\linewidth]{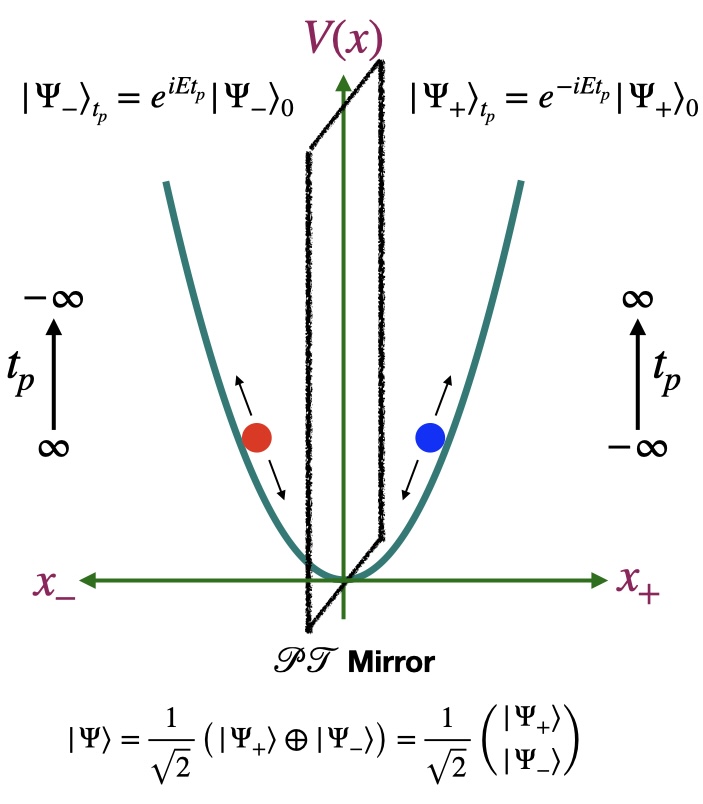}
    \caption{{In this picture, we depict the quantum harmonic oscillator in the framework of the direct-sum Schr\"{o}dinger equation, which describes a quantum state as a direct-sum of two components in the parity conjugate regions of physical space and are positive energy states in the opposite arrows of time. In this picture, the two components of the quantum state are mirror images of each other by $\Pc\Tc$. This would mean in the simple harmonic motion (quantum mechanical), if we place a ``$\Pc\Tc$ mirror’’ at $x = 0$, we would be witnessing the component quantum states $\Psi_{\pm}$ as mirror images of each other. Note that the choice of placing $\Pc\Tc$ mirror at $x=0$ is not special, in fact one can choose any non-zero origin to realize the equivalent $\Pc\Tc$ symmetric understanding. }}
    \label{fig:enter-label}
\end{figure}

We review the derivation of the wave function for the quantum harmonic oscillator in the framework of direct-sum Schr\"{o}dinger Equation \eqref{disumSch} \cite{Gaztanaga:2024vtr}. The Hamiltonian of the quantum harmonic oscillator is 
\begin{equation}
\begin{aligned}
    \hat H & =  \frac{\hat p^2}{2}+\frac{1}{2}\hat x^2 \\ 
    & = \LF\frac{\hat p_+^2}{2}+\frac{1}{2}\hat x_+^2\RF \oplus \LF\frac{\hat p_-^2}{2}+\frac{1}{2}\hat x_-^2\RF
    \end{aligned}
\end{equation}
where 
\begin{equation}
		\begin{aligned}
			\hat{x}_+ & = \frac{1}{\sqrt{2}}\LF a+a^\dagger \RF,\quad \hat{p}_+ = -i\frac{d}{dx_+}= i \frac{1}{\sqrt{2}} \LF a^\dagger-a \RF \\ 
			\hat{x}_- & = \frac{1}{\sqrt{2}}\LF b+b^\dagger \RF,\quad \hat{p}_- = i\frac{d}{dx_-}= -i\frac{1}{\sqrt{2}} \LF b^\dagger-b \RF 
		\end{aligned}
	\end{equation}
	with 
	\begin{equation}
		\begin{aligned}
			\Big[\hat{x}_+,\,\hat{p}_+\Big] & = i,\quad \Big[\hat{x}_-,\,\hat{p}_-\Big]= -i\\
			\Big[a,\,a^\dagger\Big] & =     \Big[b,\,b^\dagger\Big] =1,\quad  \Big[a,\,b^\dagger\Big]=  \Big[a,\,b\Big]=0\,. 
		\end{aligned}
	\end{equation}
{{Following} \eqref{wavefunc}, 
it is straightforward to obtain the wave function for the harmonic oscillator 
}
{\begin{equation}
		\begin{aligned}
			\Psi_n\LF x,\,t_p \RF & = \frac{1}{\sqrt{2}}\Psi_{n+}\LF t_p,\,x_+ \RF + \frac{1}{\sqrt{2}}\Psi_{n-}\LF -t_p,\,x_- \RF \\ & = 
			\frac{1}{\sqrt{2^{n+1}n!}}\LF \frac{1}{\pi} \RF^{1/4}\LT e^{-\frac{1}{2}x_+^2} H_n\LF x_+ \RF e^{-iE_n t_p}+ e^{-\frac{1}{2}x_-^2} H_n\LF x_- \RF e^{iE_n t_p}\RT
		\end{aligned}
	\end{equation}
 }{{where $H_n(z)$ is the Hermite Polynomial. We should interpret the wave functions $\Psi_{n\pm}\LF \pm t_p,\, x_{\pm} \RF$ as not superposition since they correspond to superselection sectors of total Hilbert \mbox{space \eqref{SSSHilb}.} We refer to Figure~\ref{fig:enter-label} about the description of our quantum harmonic oscillator where the orthogonal states $\vert\Psi_{\pm}\rangle $ can be seen as $\Pc\Tc$ mirror images of each other. We must note that an observer can only measure one of the states but never {simultaneously. 
The orthogonality of the wavefunctions in the global Hilbert space $\Hc$ corresponding to different energy states is given by}
 \begin{equation}
     \int_{-\infty}^{\infty} \langle \Psi_n\vert \Psi_m\rangle dx = \int_0^\infty dx_+ \langle \Psi_{n+}\vert \Psi_{m+}\rangle + \int_{-\infty}^0 dx_- \langle \Psi_{n-}\vert \Psi_{m-}\rangle dx_- = \delta_{n,m}
 \end{equation} 
{which 
follows from the orthogonal properties of Hermite polynomials.}}}

 {The expectation value of an observable is 
 \begin{equation}
     \langle \Psi \vert \Oc \vert \Psi \rangle = \frac{1}{2}\langle \Psi_+ \vert \Oc_+\vert \Psi_+\rangle + \frac{1}{2}\langle \Psi_- \vert \Oc_-\vert \Psi_-\rangle
 \end{equation}
 where $\Oc$ is an operator corresponding to an observable that splits according to our direct-sum rule (superselection by $\Pc\Tc$) as 
 \begin{equation}
     \Oc = \begin{pmatrix}
         \Oc_+ & 0 \\
         0 & \Oc_-
     \end{pmatrix}
 \end{equation}}

\section{Direct-sum quantum field theory}

\label{sec:DQFT}

According to DQFT \cite{Kumar:2023ctp,Kumar:2023hbj,Gaztanaga:2024vtr}, the Klein-Gordon (KG) field in 4D Minkowski spacetime ($ds^2 =-dt^2+d\textbf{x}^2$) is quantized as
	\begin{equation}
		\begin{aligned}
			\hat{\phi}\LF x \RF & = \frac{1}{\sqrt{2}}  \hat{\phi}_{+}  \LF t,\, \textbf{x} \RF \oplus \frac{1}{\sqrt{2}} \hat{\phi}_{-} \LF -t,\,-\textbf{x} \RF \\ 
			& = \frac{1}{\sqrt{2}} \begin{pmatrix}
				\hat{\phi}_{+} & 0 \\ 
				0 & 	\hat{\phi}_{-}
			\end{pmatrix}
		\end{aligned}
		\label{disum}
	\end{equation}
	where {$\phi_{\pm}(\pm x)$ are functions of $\Pc\Tc$ conjugate points in the Minkowski spacetime given by}
	\begin{equation}
		\begin{aligned}
			\hat{\phi}_{+}  \LF x \RF &  = 	\int \frac{d^3k}{\LF 2\pi\RF ^{3/2}}	\frac{1}{\sqrt{2\vert k_0 \vert}} \Bigg[\hat a_{(+)\,\textbf{k}}  e^{ik\cdot x}+\hat a^\dagger_{(+)\,\textbf{k}} e^{-ik\cdot x} \Bigg]   \\ 
			\hat{\phi}_{-}  \LF -x\RF &  = \int \frac{d^3k}{\LF 2\pi\RF ^{3/2}}	\frac{1}{\sqrt{2\vert k_0 \vert}} \Bigg[\hat a_{(-)\,\textbf{k}}  e^{-ik\cdot x}+\hat a^\dagger_{(-)\,\textbf{k}} e^{ik\cdot x} \Bigg] 
		\end{aligned}
		\label{fiedDQFTMin}
	\end{equation}
{where} 
$k\cdot x = -k_0t + \textbf{k}\cdot \textbf{x}$. Note that in standard QFT, we expand the field operator as
\begin{equation}
    \hat \phi(x) = \int \frac{d^3x}{\LF 2\pi \RF^{3/2}} \frac{1}{\sqrt{2\vert k_0 \vert}} \Bigg[\hat a_{\textbf{k}}  e^{ik\cdot x}+\hat a^\dagger_{\textbf{k}} e^{-ik\cdot x} \Bigg]
\end{equation}
{which} 
is dictated by Lorentz symmetries and the requirement of operators to commute for space-like distances 
\begin{equation}
  \Bigg[\hat \phi(x),\, \hat \phi(y)\Bigg]=0,\quad  \LF x-y \RF^2>0  
  \label{ccond}
\end{equation}
which is called the causality condition. The connection between QM and QFT enters by associating the positive energy state mode function $e^{-ik_0t}$ (according to the arrow of time $t:-\infty\to \infty$) to the annihilation operator $\hat a_{\textbf{k}}$. 

This construction follows straightforwardly from direct-sum representation of Schr\"{o}dinger equation 
 
	The creation and annihilation operators satisfy
	\begin{equation}
		\begin{aligned}
			[\hat{a}_{(+)\,\textbf{k}},\,\hat{a}_{(+)\,\textbf{k}^\prime}^\dagger] & = 	[\hat{a}_{(-)\,\textbf{k}},\,\hat{a}_{(-)\,\textbf{k}^\prime}^\dagger] = \delta^{(3)}\LF \textbf{k}-\textbf{k}^\prime \RF\\
			[\hat{a}_{(+)\,\textbf{k}},\,\hat{a}_{(-)\,\textbf{k}^\prime}] &=	[\hat{a}_{(+)\,\textbf{k}},\,\hat{a}_{(-)\,\textbf{k}^\prime}^\dagger]   =0\,.
		\end{aligned}
		\label{comcan}
	\end{equation}
{The above commutation relations ensure that 
\begin{equation}
    \LT \hat \phi_+,\, \hat \phi_- \RT=0
\end{equation}
which is a new causality condition in the DQFT construction along with the standard commutation of operators for space-like distances \eqref{ccond}. }

Therefore, a single quantum field of momentum $\textbf{k}$ is the direct-sum of two components evolving forward and backward in time at $\textbf{x}$ and $-\textbf{x}$. The vacuum in DQFT is defined by 
\begin{equation}
    \vert 0_M\rangle = \vert 0_{M+}\rangle \oplus \vert 0_{M-}\rangle = \begin{pmatrix}
        \vert 0_{M+}\rangle \\ \vert 0_{M-}\rangle 
    \end{pmatrix}
    \label{minvac}
\end{equation}
\begin{equation}
    \hat{a}_{(+)}\vert 0_{M+}\rangle =0,\quad \hat{a}_{(-)}\vert 0_{M-}\rangle =0
\end{equation}
{The} 
Fock space in DQFT is 
\begin{equation}
    \Fc_M = \Fc_{M+} \oplus \Fc_{M-}
\end{equation}
{The} 
single-particle states in Minkowski spacetime are described in direct-sum Hilbert space
\begin{equation}
    \Hc_M = \Hc_{M+}\oplus \Hc_{M-}
    \label{minHilbsplit}
\end{equation}

A principle here is that physics has to be observer-independent; the direct-sum QM provides a universal way to define a positive energy state independent of the arrow of time.  
The notion of an observer in classical physics differs from that of quantum physics. It is also well-known that the concept of time significantly differs in QM \cite{rovelli_2004}. Therefore, one cannot align the classical observer's clock with the notion of time in QM unless there are causal boundaries in spacetime, which we will discuss in the next sections. 

DQFT uplifts the $\Pc\Tc$ (or $\Cc\Pc\Tc$, including "charge conjugation") at every stage. Our discussion here is restricted to real scalar fields.
Now, the two-point function and propagator (related time-ordered product) have two components because of the direct-sum split of the vacuum \eqref{minvac}
\begin{equation}
\begin{aligned}
    \langle 0\vert \hat{\phi}\LF x \RF  \hat{\phi}\LF x^\prime \RF \vert 0\rangle & = \frac{1}{2} \langle 0_+\vert \hat{\phi}_+\LF x \RF  \hat{\phi}_+\LF x^\prime \RF \vert 0_+\rangle + \frac{1}{2} \langle 0_-\vert \hat{\phi}_-\LF x \RF  \hat{\phi}_-\LF x^\prime \RF \vert 0_- \rangle \\ 
     \langle 0\vert T \hat{\phi}\LF x \RF  \hat{\phi}\LF x^\prime \RF 0\vert \rangle & = \frac{1}{2} \langle 0_+\vert T\hat{\phi}_+\LF x \RF  \hat{\phi}_+\LF x^\prime \RF \vert 0_+\rangle + \frac{1}{2} \langle 0_-\vert T \hat{\phi}_-\LF x \RF  \hat{\phi}_-\LF x^\prime \RF \vert 0_- \rangle \,, 
    \end{aligned}
\end{equation}
where $T$ represents time ordering. All the interactions are also split into direct-sum components. For example, consider a cubic interaction 
\begin{equation}
  \frac{\lambda}{3} \hat{\phi}^3 = \frac{\lambda}{3} \begin{pmatrix}
      \hat{\phi}_+^3 & 0 \\ 
      0 & \hat{\phi}_-^3
  \end{pmatrix}
\end{equation}
which shows that there will never be any mixing between $\hat{\phi}_+$ and $\hat{\phi}_-$. {So, all the calculations of standard QFT can be straightforwardly extended to DQFT. However, we do not expect any change in the results of standard QFT because all the propagators and vertices of standard model degrees of freedom are split into two components. According to DQFT, the standard model vacuum and the degrees of freedom are split into two components as} 
\begin{equation}
    \vert 0_{SM}\rangle = \begin{pmatrix}
        \vert 0_{SM+}\rangle \\ 
        \vert 0_{SM-}\rangle 
    \end{pmatrix} \quad \vert SM\rangle = \frac{1}{\sqrt{2}}\begin{pmatrix}
        \vert SM_+\rangle \\ 
        \vert SM_-\rangle \end{pmatrix} \quad \vert \overline{SM}\rangle = \frac{1}{\sqrt{2}}\begin{pmatrix}
        \vert \overline{SM}_+\rangle \\ 
        \vert \overline{SM}_-\rangle 
    \end{pmatrix}
\end{equation}
{where $\vert SM\rangle$ and $\vert \overline{SM}\rangle $ represent particle and anti-particle degrees of freedom. It is important to note that every standard model degree of freedom in DQFT is represented by two components, which are applicable to both particle and anti-particle degrees of freedom. Note that we still maintain the same interpretation of anti-particles as particles going backward in time in each superselection sector. We can still describe a superposition of particle and anti-particle degrees of freedom maintained in each superselection sector }
\begin{equation}
  \alpha  \vert SM\rangle +\beta\vert \overline{SM}\rangle = \frac{1}{\sqrt{2}}\begin{pmatrix}
       \alpha \vert SM_+\rangle + \beta\vert \overline{SM}_+\rangle \\ \alpha \vert SM_-\rangle + \beta \vert \overline{SM}_-\rangle
    \end{pmatrix}
\end{equation}
{{Note} 
that if we apply DQFT for the entire standard model of particle physics, we must impose our super-selection-rule based on $\Pc\Tc$ to be the same for all Fock spaces, i.e., the parity conjugate regions are uniquely defined for all the particle and antiparticle states of the standard model. According to this, all the standard model interactions (either cubic or quartic) are split into the block diagonal form  
\begin{equation}
   \Lc_c \sim\Oc_{SM}^3=\begin{pmatrix}
        \Oc_{SM_+}^3 & 0 \\ 
        0 & \Oc_{SM_-}^3
    \end{pmatrix} \quad \Lc_q \sim \Oc_{SM}^4 = \begin{pmatrix}
        \Oc_{SM_+}^4 & 0 \\ 
        0 & \Oc_{SM_-}^4
    \end{pmatrix}
\end{equation}
{Here}, 
$\Oc_{SM}$ represents any operator in the Standard Model involving quantum fields and their derivatives. }

{DQFT is a framework that does not alter the QFT calculations in Minkowski due to the spacetime being $\Pc\Tc$ symmetric. However, extending the new understanding of spacetimes with horizons offers a resolution to the most important conundrums. This is exactly what we will learn in detail in the later sections. }

\section{DQFT in Rindler spacetime} 

\label{sec:DQFTRindler}

Rindler spacetime is a part of Minkowski spacetime in Rindler coordinates, which restricts spacetime with a boundary (horizon). The coordinate system emerges from Lorentz, the symmetries of Minkowski spacetime. 
This section particularly discusses the Rindler spacetime in (1 + 1) dimension. 

The (1 + 1) dimensional Minkowski spacetime is 
\begin{equation}
    ds^2 = -dt^2+ dz^2
\end{equation}
which is invariant under the Lorentz transformations 
\begin{equation}
    \begin{aligned}
        t &\to t\cosh{\beta}+z\sinh{\beta} \\
         z &\to t\sinh{\beta}+z\cosh{\beta}
    \end{aligned}
\end{equation}
where $\beta$ is the usual Lorentz factor. 

\begin{figure}
    \includegraphics[width=0.5\linewidth]{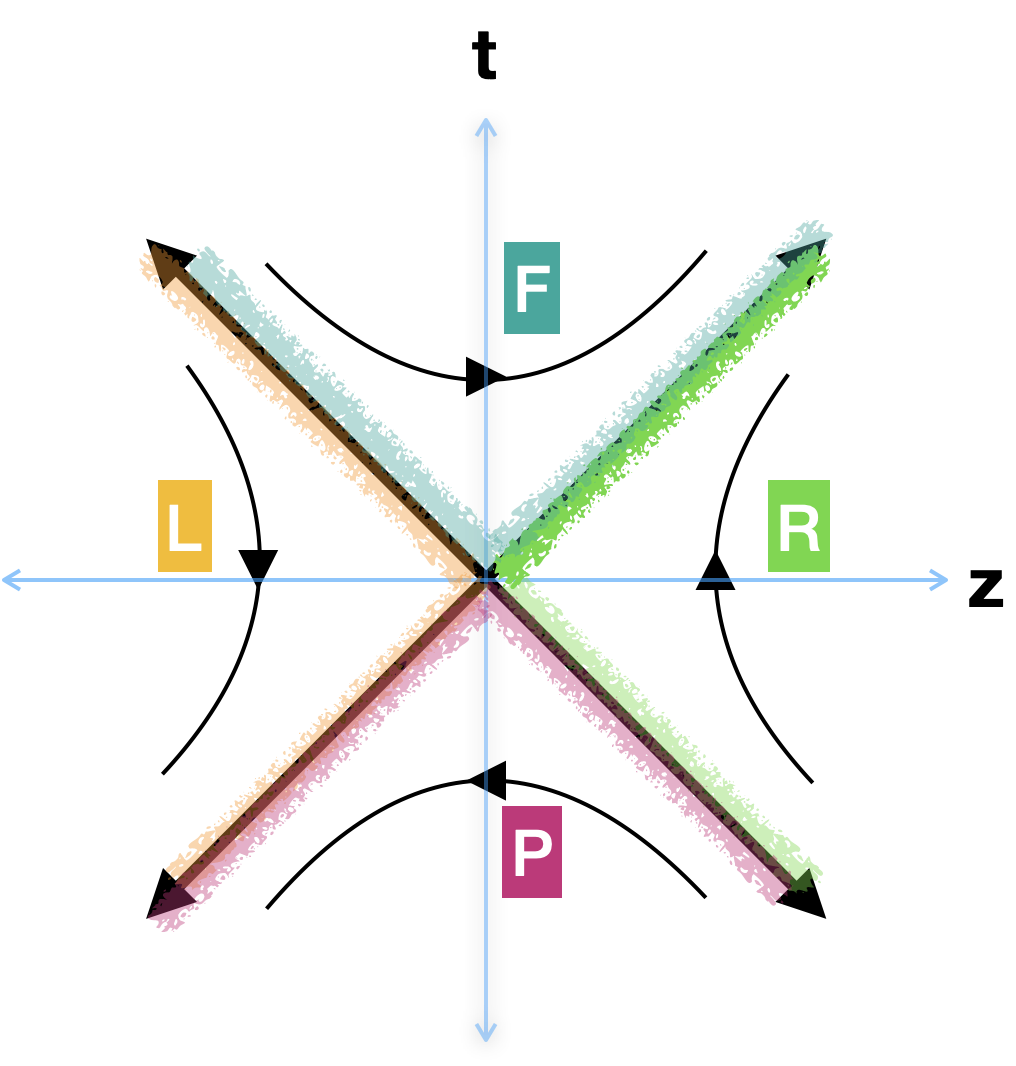}
    \caption{In this picture, we depict the Left, Right ($z^2\gtrsim t^2$) and Future Past ($t^2\gtrsim z^2$) regions of Rindler spacetime in the plane of Minkowski coordinates $\LF t,\,z \RF$. The curved lines in the Left and Right regions depict a constant acceleration $ae^{-a\xi}$ where the arrow of time in the Left goes in the direction $\eta: \infty \to -\infty$ whereas, in the Right, it is $\eta: -\infty \to \infty$. Future and Past Rindler wedges represent degenerate Kasner Universes where the arrows indicate changing $z: \mp\infty \to \pm\infty$, which means $\eta: \mp\infty \to \pm\infty$. The Fuzzy colored lines represent the Rindler Horizons for Left (Yellow), Right (Green), Future (Cyan), and Past (Pink).}
    \label{fig:RindlerST}
\end{figure}

We can now define spacetime regions that belong to the two superselection sectors of global Hilbert space that are separated by the (Rindler) horizons 
\begin{equation}
    z^2-t^2 = \frac{1}{a^2} e^{2a\xi} ,\quad t^2-z^2 = \frac{1}{a^2} e^{2a\eta}\,. 
    \label{z2t2}
\end{equation}
{From} 
\eqref{z2t2} it is obvious to define 
coordinate systems, which define the four {observers.} 
Note that in our notation $\LF\xi,\,\eta\RF$ correspond to  $\LF\eta,\, \zeta\RF$ found in \cite{Higuchi:2017gcd,Crispino:2007eb,Mukhanov:2007zz}. Furthermore, we choose to have the same coordinate symbols for Left, Right, and Future Past regions of Rindler spacetime.
\begin{equation}
\begin{aligned}
 z^2-t^2 & = \frac{1}{a^2} e^{2a\xi} \implies \begin{cases}
     z  = \frac{1}{a} e^{a\xi} \cosh{a\eta},\quad  t= \frac{1}{a} e^{a\xi} \sinh{a\eta} \quad \LF \textrm{Right Rindler} \RF \\   z  = -\frac{1}{a} e^{a\xi} \cosh{a\eta},\quad  t= \frac{1}{a} e^{a\xi}\sinh{a\eta} \quad \LF \textrm{Left Rindler} \RF
 \end{cases} \\ &\implies \boxed{ds^2 = e^{2a\xi}\LF -d\eta^2+d\xi^2\RF} \\   
  t^2-z^2 & = \frac{1}{a^2} e^{2a\eta} \implies \begin{cases}
     t  = \frac{1}{a} e^{a\eta} \cosh{a\xi},\quad  z= \frac{1}{a} e^{a\eta} \sinh{a\xi}\quad \LF \textrm{Future Kasner} \RF  \\  t  = -\frac{1}{a} e^{a\eta} \cosh{a\xi},\quad  z= \frac{1}{a} e^{a\eta} \sinh{a\xi}\quad \LF \textrm{Past Kasner} \RF 
 \end{cases} \\ & \implies \boxed{ds^2 = e^{2a\eta}\LF -d\eta^2+d\xi^2\RF}
    \end{aligned}
    \label{Robserco}
\end{equation}
{A crucial} 
observation we can make here is that 
\begin{equation}
    \begin{aligned}
        ds^2 & = e^{2a\xi}\LF -d\eta^2+d\xi^2\RF,\quad z^2-t^2\gtrsim 0 \\ 
         ds^2 & = e^{2a\eta}\LF -d\eta^2+d\xi^2\RF,\quad t^2-z^2\gtrsim 0\,.  
    \end{aligned}
    \label{Robser}
\end{equation}
{Here} 
$ae^{-a\xi}$ is the proper acceleration in the Left and Right Rindler observer in the direction of the respective Killing vectors ($\frac{\pd}{\pd \eta}$) that define the arrow of time $t(\eta)$ for each observer.
The Future and Past Kasner spacetime corresponds to expanding Universes in opposite arrows of time as depicted in Figure~\ref{fig:RindlerST}. 
The equality 
$z^2= t^2$ (i.e., in the limit $ae^{-a\xi}\to \infty$) defines all the Rindler horizons corresponding to Left, Right Rindler, Future, and Past Kasner observers \cite{Higuchi:2017gcd} (See Figure~\ref{fig:RindlerST}). Another crucial observation to make here is that the nature of spacetime gets interchanged (i.e., $z\longleftrightarrow t$ together with $\xi \longleftrightarrow$ $\eta$), going from Right to Future and Left to Past Rindler regions. 

We can rewrite the whole Rindler spacetime (with all regions) in a coordinate system defined by 
\begin{equation}
    \begin{aligned}
        U &= -\frac{1}{a}e^{-au}<0,\quad &&V= \frac{1}{a}e^{av}>0\quad &&(\rm Right\, Rindler)\\
        U&= \frac{1}{a}e^{-au}>0,\quad &&V= -\frac{1}{a}e^{av}<0\quad &&(\rm Left\, Rindler) \\
         U&= \frac{1}{a}e^{-au}>0,\quad &&V= \frac{1}{a}e^{av}>0\quad &&(\rm Future\, Kasner) \\
          U&= -\frac{1}{a}e^{-au}<0,\quad &&V= -\frac{1}{a}e^{av}<0\quad &&(\rm Past\, Kasner)
    \end{aligned}
    \label{UVcoord}
\end{equation}
where 
\begin{equation}
\begin{aligned}
    u&= \eta-\xi,\quad v=\eta+\xi \\ 
    U&= t-z,\quad V=t+z
    \end{aligned}
\end{equation}
{In this} 
coordinates \eqref{UVcoord}, the transition between Left, Right, Future, and Past Rindler observers is carried through discrete transformations on $\LF U,\, V \RF$. 
In other words, crossing horizons is nothing but applying discrete operations on the coordinates $\LF U,\, V \RF$.
This is analogous to the context of so-called Kruskal coordinates in black hole spacetime, \cite{Kumar:2023hbj}, which we shall return to in the later section. 

\subsection{DQFT in Right and Left Rindler Spacetime}

Given this configuration, we first derive how DQFT vacuum \eqref{minvac} would look to the Left and Right Rindler regions where the arrow of time (the direction of the uniformly accelerated observer) in each other's is opposite. 

According to DQFT, we split the KG field operator into the direct-sum components corresponding to parity conjugate regions of physical spacetime. Thus, in the \mbox{regions $z>\vert t\vert $,} we can write the KG operator in Minkowski and Rindler spacetimes as 
\begin{equation}
\begin{aligned}
    \hat{\phi} & = \frac{1}{\sqrt{2}} \LF \hat{\phi}_+\oplus \hat{\phi}_-  \RF = \frac{1}{\sqrt{2}}\begin{pmatrix}
        \hat{\phi}_+ & 0 \\
        0 & \hat{\phi}_- 
    \end{pmatrix}\Bigg\vert_{z^2\gtrsim t^2\, {\rm Minkowski}} \\ &  =    \frac{1}{\sqrt{2}}\LF \hat{\phi}_R\oplus \hat{\phi}_L \RF = \frac{1}{\sqrt{2}} \begin{pmatrix}
        \hat{\phi}_{R} & 0 \\
        0 & \hat{\phi}_{L} 
    \end{pmatrix} \,,
    \label{eqLRM}
    \end{aligned}
\end{equation}
where the subscripts $L,\, R$ represent field operators expressed in the Left Rindler and the Right Rindler coordinates, respectively. 
From the definitions of the Right and Left Rindler coordinates \eqref{Robserco} we can notice that they cover the parity conjugate regions $z$ and $-z$ with arrows of time $\eta: -\infty \to \infty$ (which imply $t: -\infty \to \infty $) and $\eta: \infty \to -\infty $  (which imply $t: \infty \to -\infty $) respectively. {We can clearly notice here that the Left and Right Rindler wedges are $\Pc\Tc$ mirror of each other. Since we express our field operators in Minkowski as direct-sum two components which are $\Pc\Tc$ mirrors of each other, we thus map $\hat \phi_{+}\to \hat \phi_R$ and $\hat \phi_{-}\to \hat \phi_{L}$ and evaluate the respective Bogoliubov coefficients. This is because the arrow of time in the Right Rindler wedge $\eta: -\infty \to \infty$ coincides with the arrow of time in the Minkowski vacuum $\vert 0_{M_+}\rangle $ \eqref{minvac}. Similarly, the arrow of time in the Left Rindler wedge coincides with the arrow of time in the Minkowski vacuum $\vert 0_{M-}\rangle$ \eqref{minvac}. Remember that in our construction, flipping the arrow of time is associated with going to a parity conjugate region. We construct local Hilbert space or Fock space associated with parity conjugate regions of physical space. This is the reason why we decompose the field operators as written in \eqref{eqLRM}. }

Following \eqref{eqLRM} we obtain 
\begin{equation}
\begin{aligned}
    \hat{\phi}_+ & = \int \frac{dk}{(2\pi)^{1/2}}\frac{1}{\sqrt{2\vert k \vert}} \Bigg[\hat c_{(+)\,\textbf{k}}  e^{-ikt+ikz}+\hat c^\dagger_{(+)\,\textbf{k}} e^{ikt-ikz} \Bigg] \\ 
&  =   \int \frac{dp}{(2\pi)^{1/2}}\frac{1}{\sqrt{2\vert p \vert}} \Bigg[\hat c_{R\,\textbf{p}}  e^{-ip\eta+ip\xi}+\hat c^\dagger_{R\,\textbf{p}} e^{ip\eta-ip\xi} \Bigg] \\
   \hat{\phi}_- & = \int \frac{dk}{(2\pi)^{1/2}}\frac{1}{\sqrt{2\vert k \vert}} \Bigg[\hat c_{(-)\,\textbf{k}}  e^{ikt-ikz}+\hat a^\dagger_{(-)\,\textbf{k}} e^{-ikt+ikz} \Bigg] \\ 
&  =   \int \frac{dp}{(2\pi)^{1/2}}\frac{1}{\sqrt{2\vert p \vert}} \Bigg[\hat c_{L\,\textbf{p}}  e^{ip\eta-ip\xi}+\hat c^\dagger_{L\,\textbf{p}} e^{-ip\eta+ip\xi} \Bigg]
    \end{aligned} 
\end{equation}
{Here} 
$\LF \hat c_{\pm},\,\hat c^\dagger_{\pm} \RF$ satisfy the canonical commutation relations similar to \eqref{comcan}. All the operators corresponding to the Fock space of the Left Rindler commute with the Fock space operators of the Right Rindler spacetime. 

Recall that the Minkowski vacuum is 
\begin{equation}
    \hat c_\textbf{k}\vert 0\rangle = \begin{pmatrix}
        \hat c_{(+)\,\textbf{k}} & 0 \\ 
        0 & \hat c_{(-)\,\textbf{k}} 
    \end{pmatrix}\begin{pmatrix}
        \vert 0_+\rangle \\ 
        \vert 0_-\rangle 
    \end{pmatrix} = 0\implies \langle 0 \vert \hat{c}_\textbf{k}^\dagger \hat{c}_\textbf{k} \vert 0\rangle =0\,. 
\end{equation}
whereas the Rindler vacuum is
\begin{equation}
    \hat{c}_{A\,\textbf{p}}\vert 0_A\rangle = \begin{pmatrix}
        \hat{c}_{L\,\textbf{p}} & 0 \\ 
        0 & \hat{c}_{R\,\textbf{p}}
    \end{pmatrix} \begin{pmatrix}
        \vert 0_L\rangle \\ 
        \vert 0_R\rangle 
    \end{pmatrix} = 0\,. 
\end{equation}
{Here}, 
the subscript $A$ denotes the Left and Right Rindler spacetime spanned by coordinates satisfying $z^2-t^2 \gtrsim 0$. 
Following Appendix~\ref{appbogo} we can deduce 
\begin{equation}
    \begin{pmatrix}
        \hat{c}_{R \textbf{p}} \\ \hat{c}_{R\textbf{p}}^\dagger  
    \end{pmatrix} = \int dk \begin{pmatrix}
        \alpha^R_{kp} & \beta^R_{kp} \\ 
        \beta^{R\ast}_{kp} & \alpha_{kp}^{R\ast}
    \end{pmatrix} \begin{pmatrix}
        \hat{c}_{(+)\,\textbf{k}} \\ \hat{c}^\dagger_{(+)\,\textbf{k}}
    \end{pmatrix}
    \label{CAbogR}
\end{equation}
and 
\begin{equation}
    \begin{pmatrix}
        \hat{c}_{L \textbf{p}} \\ \hat{c}_{L\textbf{p}}^\dagger  
    \end{pmatrix} = \int dk \begin{pmatrix}
        \alpha^L_{kp} & \beta^L_{kp} \\ 
        \beta^{L\ast}_{kp} & \alpha_{kp}^{L\ast}
    \end{pmatrix} \begin{pmatrix}
        \hat{c}_{(-)\,\textbf{k}} \\ \hat{c}^\dagger_{(-)\,\textbf{k}}
    \end{pmatrix}
    \label{CAbogL}
\end{equation}
{The} 
above relations imply the particle number density for each Rindler observer in the Minkowski vacuum is non-zero. This means 
\begin{equation}
\begin{aligned}
 \langle N_A\rangle  = \langle 0 \vert \hat c_{A\,\textbf{p}}^\dagger  \hat c_{A\,\textbf{p}} \vert 0 \rangle & =   \langle 0_-\vert \hat c_{R\,\textbf{p}}^\dagger\hat c_{R\,\textbf{p}} \vert 0_- \rangle \Theta(z) +\langle 0_+\vert \hat c_{L\,\textbf{p}}^\dagger\hat c_{L\,\textbf{p}} \vert 0_+ \rangle \Theta(-z) \\ 
    & = \Theta(z)\int dk \vert \beta_{kp}^R\vert^2 +\Theta(-z)\int dk \vert \beta_{kp}^L\vert^2\,.
    \end{aligned}
\end{equation}
{It is} 
important to keep in mind that a single state in Rindler spacetime satisfying $z^2-t^2 \gtrsim 0$ is the direct-sum of two components 
\begin{equation}
    \vert \phi_A  \rangle = \frac{1}{\sqrt{2}}\LF {\vert \phi_L\rangle \oplus \vert \phi_R \rangle } \RF = \frac{1}{\sqrt{2}}\begin{pmatrix}
    \vert \phi_L \rangle \\ \vert \phi_R \rangle
    \end{pmatrix}
\end{equation}
where $\vert \phi_A\rangle = \hat{\phi}_A \vert 0_A\rangle $, $\vert \phi_L\rangle = \hat{\phi}_L\vert 0_L\rangle $ and $\vert \phi_R\rangle = \hat{\phi}_R \vert 0_R\rangle $. 
The two components states $\vert \phi_{L/R}\rangle $ belong to superselection sectors of total Hilbert space where a single state $\vert \phi_A\rangle$ is defined. Thus, the Hilbert space of a single particle state becomes 
\begin{equation}
    \Hc_A = \Hc_L \oplus \Hc_R\,,
    \label{LRhilber}
\end{equation}
whereas the Fock space of QFT in Rindler spacetime (that covers $z^2\gtrsim t^2$) becomes 
\begin{equation}
    \Fc_A = \Fc_L \oplus \Fc_R 
\end{equation}
{Once} 
more, recall that the arrow of time is not the same in the Left and Right Rindler spacetime. We can see it as a thumb rule to apply the superselection rule for Hilbert space whenever we see a change in the arrow of time in the target space. 

{The Bogoliubov coefficients can be can be calculated as follows using the mathematical tequniques described in \cite{Mukhanov:2007zz} }

\begin{equation}
\begin{aligned}
    \alpha^R_{kp} & = \sqrt{\frac{\Omega }{\omega }} B\LF \omega,\,\Omega \RF, \quad \alpha^L_{kp}  = \sqrt{\frac{\Omega }{\omega }} B^\ast\LF \omega,\,\Omega \RF \\ 
   \beta^R_{kp} & = \sqrt{\frac{\Omega }{\omega }} B\LF -\omega,\,\Omega \RF, \quad  \beta^L_{kp}  = \sqrt{\frac{\Omega }{\omega }} B^\ast\LF -\omega,\,\Omega \RF 
    \end{aligned}
\end{equation}
where $\omega = \vert k \vert $, $\Omega = \vert p\vert $ and 
\begin{equation}
    B\LF \omega,\,\Omega \RF = \frac{1}{2\pi a} exp\LT \frac{i\Omega}{a}\ln\Big\vert \frac{\omega}{a} \Big\vert +\frac{\pi\Omega}{2a}{\rm sign}\LF \frac{\omega}{a} \RF\RT \Gamma\LF -\frac{i\Omega}{a} \RF 
\end{equation}

{The mean density of particles that Left and Right Rindler observers perceive is obtained as}

\begin{equation}
\begin{aligned}
    n_{\Omega}^L & = \int_0^\infty d\omega \Big\vert  \beta_{kp}^L \Big\vert^2 = \Big[ {\rm exp}\LF \frac{2\pi\Omega}{a} \RF-1\Big]^{-1} \\ 
    n_{\Omega}^R & = \int_0^\infty d\omega \Big\vert  \beta_{kp}^R\Big\vert^2 = \Big[ {\rm exp}\LF \frac{2\pi\Omega}{a} \RF-1\Big]^{-1} \\
    \end{aligned}
    \label{NpLR}
\end{equation}
{Thus}, 
comparing the above expressions with the Bose-Einstein distribution 
\begin{equation}
n(\Omega) = \Big[{\rm exp}\LF \frac{E}{T}  \RF-1\Bigg]^{-1}
\end{equation}
we get the temperature associated with these particles as 
\begin{equation}
    T_R = \frac{a}{2\pi}
\end{equation}

\subsection{DQFT in Past and Future Rindler Spacetime}

We can easily deduce from \eqref{Robserco} that the definitions of time and space get swapped from Right or Left ($ z^2\gtrsim t^2 $) to Future or Past ($ t^2\gtrsim z^2 $) Rindler spacetimes. In the Future Rindler wedge, $z: -\infty \to \infty$ represents the direction of the Killing vector. Whereas in the Past Rindler Wedge, we have $z: \infty \to -\infty$ denoting the direction of the Killing \mbox{vector $z\frac{\pd}{\pd t}+t\frac{\pd}{\pd z}$. }

{Thus, according to  Figure~\ref{fig:RindlerST}}
we expand the scalar field operator in Minkowski, Future and Past Rindler spacetimes as 
\begin{equation}
\begin{aligned}
    \hat{\phi} & = \frac{1}{\sqrt{2}} \LF \hat{\phi}_+\oplus \hat{\phi}_-  \RF = \frac{1}{\sqrt{2}}\begin{pmatrix}
        \hat{\phi}_+ & 0 \\
        0 & \hat{\phi}_- 
    \end{pmatrix}\Bigg\vert_{t\gtrsim \vert z\vert \, {\rm Minkowski}} \\ &  =   \frac{1}{\sqrt{2}} \LF \hat{\phi}_{F+}\oplus \hat{\phi}_{F-} \RF = \frac{1}{\sqrt{2}}\begin{pmatrix}
        \hat{\phi}_{F+} & 0 \\
        0 & \hat{\phi}_{F-} 
    \end{pmatrix} 
    \label{eqFP}
    \end{aligned}
\end{equation}
and 
\begin{equation}
\begin{aligned}
    \hat{\phi} & = \frac{1}{\sqrt{2}} \LF \hat{\phi}_+\oplus \hat{\phi}_-  \RF = \frac{1}{\sqrt{2}}\begin{pmatrix}
        \hat{\phi}_+ & 0 \\
        0 & \hat{\phi}_- 
    \end{pmatrix}\Bigg\vert_{t\lesssim -\vert z\vert \, {\rm Minkowski}} \\ &  =   \frac{1}{\sqrt{2}} \LF \hat{\phi}_{P+}\oplus \hat{\phi}_{P-} \RF = \frac{1}{\sqrt{2}}\begin{pmatrix}
        \hat{\phi}_{P+} & 0 \\
        0 & \hat{\phi}_{P-} 
    \end{pmatrix}
    \label{eqFP2}
    \end{aligned}
\end{equation}
where 
\begin{equation}
\begin{aligned}
    \hat{\phi}_+ & = \int \frac{dk}{(2\pi)^{1/2}}\frac{1}{\sqrt{2\vert k\vert }} \Bigg[\hat c_{(+)\,\textbf{k}}  e^{-ik t+ikz}+\hat c^\dagger_{(+)\,\textbf{k}} e^{ik t-ikz} \Bigg] \\ 
&  =   \int \frac{dp}{(2\pi)^{1/2}}\frac{1}{\sqrt{2\vert p \vert}} \Bigg[\hat c_{(K+)\,\textbf{p}}  e^{-ip\eta+ip\xi}+\hat c^\dagger_{(K+)\,\textbf{p}} e^{ip\eta-ip\xi} \Bigg] =\hat{\phi}_{K+} \\
   \hat{\phi}_- & = \int \frac{dk}{(2\pi)^{1/2}}\frac{1}{\sqrt{2\vert k \vert}} \Bigg[\hat c_{(-)\,\textbf{k}}  e^{ikt-ikz}+\hat c^\dagger_{(-)\,\textbf{k}} e^{-ikt+ikz} \Bigg] \\ 
&  =   \int \frac{dp}{(2\pi)^{1/2}}\frac{1}{\sqrt{2\vert p \vert}} \Bigg[\hat c_{(K-)\,\textbf{p}}  e^{ip\eta-ip\xi}+\hat c^\dagger_{(K-)\,\textbf{p}} e^{-ip\eta+ip\xi} \Bigg] = \hat{\phi}_{K-}
    \end{aligned} 
\end{equation}
{The} 
Bogoliubov transformations are given by 

\begin{equation}
    \begin{pmatrix}
        \hat{c}_{(K+) \textbf{p}} \\ \hat{c}_{(K+)\textbf{p}}^\dagger  
    \end{pmatrix} = \int dk \begin{pmatrix}
        \alpha^{K+}_{kp} & \beta^{K+}_{kp} \\ 
        \beta^{{K+}\ast}_{kp} & \alpha_{kp}^{{K+}\ast}
    \end{pmatrix} \begin{pmatrix}
        \hat{c}_{(+)\,\textbf{k}} \\ \hat{c}^\dagger_{(+)\,\textbf{k}}
    \end{pmatrix}
    \label{FAbogF}
\end{equation}
and 
\begin{equation}
    \begin{pmatrix}
        \hat{c}_{(K-) \textbf{p}} \\ \hat{c}_{(K-)\textbf{p}}^\dagger  
    \end{pmatrix} = \int dk \begin{pmatrix}
        \alpha^{K-}_{kp} & \beta^{K-}_{kp} \\ 
        \beta^{K-\ast}_{kp} & \alpha_{kp}^{K-\ast}
    \end{pmatrix} \begin{pmatrix}
        \hat{c}_{(-)\,\textbf{k}} \\ \hat{c}^\dagger_{(-)\,\textbf{k}}
    \end{pmatrix}
    \label{FABog}
\end{equation}
where $K= F$ for $t\gtrsim \vert z\vert$ and $K=P$ for $t\lesssim -\vert z\vert$. Here $\LF \hat c_{(K\pm)},\,\hat c^\dagger_{(K\pm)} \RF$ satisfy the commutation relations similar to \eqref{comcan}. All the operators corresponding to the Fock space of the Future Rindler commute with the Fock space operators of the Past Rindler spacetime.

The Bogoliubov coefficients can be computed straightforwardly using the standard methods described in \cite{Mukhanov:2007zz}, and they are given by
\begin{equation}
\begin{aligned}
    \alpha^{K+}_{kp} & = \sqrt{\frac{\Omega }{\omega }} B^\ast\LF \omega,\,\Omega \RF, \quad \alpha^{K-}_{kp}  = \sqrt{\frac{\Omega }{\omega }} B\LF \omega,\,\Omega \RF \\ 
   \beta^{K+}_{kp} & = \sqrt{\frac{\Omega }{\omega }} B^\ast\LF -\omega,\,\Omega \RF, \quad  \beta^{K-}_{kp}  = \sqrt{\frac{\Omega }{\omega }} B\LF -\omega,\,\Omega \RF 
    \end{aligned}
\end{equation}
{From} 
the above expansion of the field operator in the Future and Past regions, we can deduce that the respective Hilbert spaces are split by
\begin{equation}
\begin{aligned}
    \Hc_F = \Hc_{F+}\oplus \Hc_{F-},\quad    \Hc_P = \Hc_{P+}\oplus \Hc_{P-} 
    \end{aligned}
\end{equation}
which is analogous to the direct-sum split of Hilbert space in Minkowski spacetime \eqref{minHilbsplit}. 
The mean density of particles in the Future and Past Rindler spacetime can be obtained in analogous way to \eqref{NpLR} as below
\begin{equation}
\begin{aligned}
    n_{\Omega}^K & = \frac{1}{2}\int_0^\infty d\omega \Big[\Big\vert  \beta_{kp}^{K+} \Big\vert^2+\Big\vert  \beta_{kp}^{K-} \Big\vert^2\Big] = \Big[ {\rm exp}\LF \frac{2\pi\Omega}{a} \RF-1\Big]^{-1} 
    \end{aligned}
    \label{NpFP}
\end{equation}

Since the Future and Past Kasner regions (See Figure~\ref{fig:RindlerST}) are separated by horizons, and the Killing vectors are in opposite directions, we must apply the superselection rule and the Hilbert spaces $\Hc_F,\, \Hc_P$ are the superselection sectors of total Hilbert space of the regions $t^2\gtrsim z^2$. Whenever we encounter Horizons, the concept of time changes on either side of the Horizons; thus, one must combine Hilbert spaces with superselection rules. 
One can, in principle, extend this concept to dynamical spacetimes, but here, our discussion is restricted to Rindler spacetime. The picture is analogous to QFT in the context of black holes and de Sitter spacetimes, too \cite{Kumar:2023ctp,Kumar:2023hbj}. 
We can notice that the Left and Right Rindler regions are separated by horizons with Future and Past Rindler regions (See Figure~\ref{fig:RindlerST}). The total Hilbert space of the entire Rindler spacetime is 
\begin{equation}
\begin{aligned}
    \Hc_{\rm total} & = \LF \Hc_L\oplus \Hc_R\RF \oplus \LF \Hc_F\oplus \Hc_P \RF\,.
    \end{aligned}
\end{equation}
{Once} 
{more, the thumb rule to construct superselection sectors of Hilbert space is when we encounter regions of spacetime related by discrete transformations, which we can see from \eqref{UVcoord}. }

\section{Entanglement and Reeh-Shlieder theorem in DQFT}

\label{sec:RStheoremDQFT}

Reeh-Schlieder (RS) theorem \cite{Reeh:1961ujh} is an important element of QFT, which tells us about the vacuum structure of Minkowski spacetime \cite{Witten:2018zxz,Agullo:2023fnp}. The theorem tells us that one can prepare a set of local operators in a sub-region  and act on the vacuum (given by \eqref{RSstate})  to approximate any arbitrary state in the total Hilbert space of the Minkowski spacetime
\begin{equation}
  \vert \Psi\rangle_{\rm RS} =   {\hat \phi}_1(x_1){\hat \phi}_2(x_1)\cdots {\hat \phi}_n(x_n)\vert 0\rangle
  \label{RSstate}
\end{equation}
where $x_i$ denote the spacetime in a sub-region and $\hat{\phi}_i(x_i)$ is a local operator in that region. This means that any sub-region in Minkowski spacetime is highly dense, and any two sub-regions of Minkowski spacetime are highly entangled. 

The proof of the RS theorem can be found in \cite{Haag:1992hx}, which we state here briefly. If the theorem is wrong, then one must find a state $\chi$ orthogonal to the state \eqref{RSstate} in the total Hilbert space of Minkowski spacetime, i.e., 
\begin{equation}
    \langle  \chi \vert \Psi\rangle_{\rm RS} =0\,.
    \label{zeroRS}
\end{equation}
{However}, 
\eqref{zeroRS} can be satisfied if the Hilbert space of Minkowski spacetime contains superselection sectors \cite{Witten:2018zxz}. The vectors in superselection sectors are orthogonal, and there cannot be any superposition between their states. 

In the DQFT, Hilbert space of Minkowski spacetime is split into superselection sectors by the parity conjugate regions of physical space (See \eqref{minHilbsplit}).  
Also, the vacuum of Minkowski spacetime is split into direct-sum components as in \eqref{minvac}. Now, if construct a state with a set of local operators acting on vacuum $\vert 0_+\rangle$ 
\begin{equation}
 \vert \Psi\rangle_{\rm RS} =    {\hat \phi}_{1+}(x_1){\hat \phi}_{2+}(x_1)\cdots {\hat \phi}_{n+}(x_n)\vert 0_+\rangle
\end{equation}
has an orthogonal state in the Hilbert space $\Hc_M$ given by 
\begin{equation}
    \vert \chi_-\rangle = \hat{\chi}_{-}\vert 0_-\rangle,\quad \langle \chi_-\vert \Psi\rangle_{\rm RS} = 0\,. 
\end{equation}
{This} 
may seem to disprove the RS theorem, but it is not surprising because not having superselection sectors is a pre-condition for the applicability of the RS theorem \cite{Witten:2018zxz}. Thus, the RS theorem is still valid separately in $\Hc_{M+}$ and $\Hc_{M-}$, the superselection sectors of total Hilbert space \eqref{minHilbsplit}. Further meaning of this to the entanglement in Minkowski spacetime is discussed in the following sub-section. 

\section{Entanglement in DQFT and Rindler spacetime: Thermal radiation and purity}

\label{sec:EqntanglePure}

The RS theorem in DQFT means that any entanglement of sub-regions in Minkowski spacetime has $\Pc\Tc$ components. 
Since parity conjugate regions of physical space are superselection sectors, any entanglement in spacetime is split into two \mbox{superselection sectors.} 

This means the following: let us consider a pure (maximally entangled) state $\vert \psi_{12}\rangle =  \vert \phi_1\rangle \otimes \vert\phi_2\rangle \in \Hc_1 \otimes \Hc_2$ which means square of the density matrix satisfies the property \mbox{of idempotence}
\begin{equation}
    \rho = \vert \psi_{12} \rangle\langle \psi_{12}\vert,\quad \rho_{12}^2=\rho_{12}
\end{equation}
{But} 
the reduced density matrix of the states $\vert \phi_1\rangle$ and $\vert \phi_2\rangle $ are not idempotent 
\begin{equation}
    \rho_1 = {\rm Tr}_{2}\rho_{12}  \neq \rho_1^2,\quad    \rho_2 = {\rm Tr}_{1}\rho_{12}  \neq \rho_2^2\,,
\end{equation}
because of which the states $\vert \phi_1\rangle $ and $\vert \phi_2\rangle$ are the mixed states of Hilbert spaces $\Hc_1$ and $\Hc_2$ respectively. This means the state $\vert \psi_{12}\rangle $ is non-factorizable, i.e.,
\begin{equation}
    \vert \psi_{12}\rangle =  \sum_{m,n} c_{mn} \vert \phi_{m1} \rangle \otimes \vert \phi_{n2}\rangle 
\end{equation}
where $c_{mn}\neq c_m c_n$ with $\vert \phi_1\rangle = \sum_m c_m \vert \phi_{m1}\rangle $ and $\vert \phi_2\rangle = \sum_n c_n \vert \phi_{n2}\rangle $.
Since every state in DQFT is direct-sum of the two components, we have

\begin{equation}
\begin{aligned}
    \vert \psi_{12}\rangle & = \sum_{m,n} c_{mn} \LF \vert \phi_{m1+} \rangle \otimes \vert \phi_{n2+}\rangle \RF \oplus \sum_{m,n} c_{mn} \LF \vert \phi_{m1-} \rangle \otimes \vert \phi_{n2-}\rangle \RF \\ & = \begin{pmatrix}
        \sum c_{mn}\vert \phi_{m1+} \rangle \otimes \vert \phi_{n2+} \rangle \\ 
        \sum c_{mn}\vert \phi_{m1-} \rangle \otimes \vert \phi_{n2-} \rangle
    \end{pmatrix} =  \begin{pmatrix}
        \vert \psi_{12+}\rangle  \\ 
         \vert \psi_{12-} \rangle 
    \end{pmatrix}
    \end{aligned}
    \label{entDQFT}
\end{equation}
{The} 
density matrix of pure state $\vert \psi_{12}\rangle =\vert \phi_1\rangle \otimes \vert \phi_2\rangle$ becomes a direct-sum of two pure state density matrices 
\begin{equation}
    \rho = \rho_+\oplus \rho_- 
\end{equation}
where $\rho_{\pm} = \vert \psi_{12\pm}\rangle \langle \psi_{12\pm}\vert $ for which Von Neumann entropies vanish $S_\pm = -{\rm Tr}\rho_\pm \ln\rho_\pm = 0$. Let us apply a similar scheme to Left/Right Rindler spacetime, where a quantum state is expressed as direct-sum of Left and Right Rindler spacetime (defined by $z^2\gtrsim t^2$) belonging to the Hilbert space \eqref{LRhilber}. 

Let us consider a (maximally) entangled pure state in Rindler spacetime ($z^2-t^2\gtrsim 0$) $\vert \psi_{LR}\rangle = \vert \phi_{A1}\rangle \otimes \vert \phi_{A2}\rangle \in \Hc_{A1}\otimes \Hc_{A2}$ which according to \eqref{entDQFT} becomes a direct-sum of two pure state components in Left Rindler and Right Rindler spacetime. 
\begin{equation}
    \vert \psi_{LR}\rangle =\begin{pmatrix}
        \vert \phi_{R1} \rangle \otimes \vert \phi_{R2} \rangle \\ 
        \vert \phi_{L1} \rangle \otimes \vert \phi_{L2} \rangle
    \end{pmatrix}
\end{equation}
{The} 
Left and Right Rindler spacetimes are separated by Horizons. According to DQFT in Rindler spacetime, the horizon separates a pure state into two pure state components of the superselection sectors, each accessed by observers on either side. 

Most importantly, any observer can reconstruct the state behind the horizon by accessing the pure state component within his/her horizon. 
This means the Right observer accesses the pure state $\vert \phi_{R1} \rangle \otimes \vert \phi_{R2} \rangle$, entanglement between the components that evolve forward in time. In contrast, the pure state beyond the Horizon of Right Rindler spacetime is nothing but entanglement between the component states evolving backward in time ($\vert \phi_{L1} \rangle \otimes \vert \phi_{L2} \rangle$) in the Left Rindler region. This implies that the Right and Left Rindler observers share complementary information in the form of pure states. In other words, both Left and Right observers do not have any unitarity loss but, at the same time, can reconstruct information beyond the horizon by observing the complete set of states within the horizon. This is exactly what we can call entanglement with purity. 



We can build a similar outcome when we include the Future and Past Kasner parts of Rindler spacetime (See Figure~\ref{fig:RindlerST}). A quantum state in the entire Rindler spacetime is a direct-sum of 4 components. 

\begin{equation}
    \vert \phi\rangle\Big\vert_{\rm All\,Rindler} = \LF\vert \phi_L\rangle\oplus \vert \phi_R\rangle \RF \oplus \LF\vert \phi_F\rangle\oplus \vert \phi_P\rangle \RF
    \label{splitall}
\end{equation}
{This} 
is because the relation between different regions of Rindler spacetime is discrete transformations \eqref{UVcoord}, and all 4 regions are separated by horizons. Thus, all 4 regions must build superselection sectors for Hilbert or Fock space. From \eqref{splitall}, it is straightforward to see how an entangled pair (pure state) can split into 4-component entangled pairs, which are again the pure states of sectorial Hilbert spaces. 

A natural question occurs here: what happens when a state evolves from one superselection sector to another superselection sector? First of all, such an evolution has to be described by discrete operators \eqref{UVcoord} on the full state $\vert \phi\rangle\Big\vert_{\rm All\,Rindler}$ and when we do that when a component evolves from Right to Future, we by construction allow a complementary state from Past to Right. This means a flow of information that keeps the pure states pure in all regions of Rindler spacetime. This happens thanks to our direct-sum split of Hilbert space by discrete spacetime operations. 

Thus, from \eqref{VACcom}, the Minkowski vacuum of DQFT (for the regions $z^2\gtrsim t^2$) can be expressed as excitations of Rindler vacuums as 
\begin{equation}
    \vert 0_M\rangle = \begin{pmatrix}
        \vert 0_{M+}\rangle  \\ 
        \vert 0_{M-}\rangle 
    \end{pmatrix} =  \begin{pmatrix}
      \prod_{\textbf{p}} \frac{1}{\sqrt{\vert \alpha_{kp}^R}\vert} \exp\Bigg[-\LF\frac{\beta^R_{kp}}{2\alpha^R_{kp}}\RF \hat{c}_{R\textbf{p}}^\dagger \hat{c}^\dagger_{R(-\textbf{p})}\Bigg] \vert 0_R\rangle \\
     \prod_{\textbf{p}} \frac{1}{\sqrt{\vert \alpha^L_{kp}}\vert} \exp\Bigg[-\LF\frac{\beta^L_{kp}}{2\alpha^L_{kp}}\RF \hat{c}_{L\textbf{p}}^\dagger \hat{c}^\dagger_{L(-\textbf{p})}\Bigg] \vert 0_L\rangle
    \end{pmatrix}
\end{equation}
{This} 
means the entangled (component) pairs (of opposite 3 momenta) in the Left and Right are related by direct-sum (because horizon or discrete transformation \eqref{UVcoord} separates them). 
Similarly, for $t^2\gtrsim z^2$ we obtain
 \begin{equation}
    \vert 0_M\rangle = \begin{pmatrix}
        \vert 0_{M+}\rangle  \\ 
        \vert 0_{M-}\rangle 
    \end{pmatrix} =   \begin{pmatrix}
       \prod_{\textbf{p}}\frac{1}{\sqrt{\vert \alpha_{kp}^{K+}}\vert} \exp\Bigg[-\LF\frac{\beta^{K+}_{kp}}{2\alpha^{K+}_{kp}}\RF \hat{c}_{(K+)\textbf{p}}^\dagger \hat{c}^\dagger_{{(K+)}\,(-\textbf{p})}\Bigg] \vert 0_{K+}\rangle \\
      \prod_{\textbf{p}}\frac{1}{\sqrt{\vert \alpha^{K-}_{kp}}\vert} \exp\Bigg[-\LF\frac{\beta^{K-}_{kp}}{2\alpha^{K-}_{kp}}\RF \hat{c}_{({K-})\textbf{p}}^\dagger \hat{c}^\dagger_{({K-})(-\textbf{p})}\Bigg] \vert 0_{K-}\rangle\
    \end{pmatrix}
\end{equation}

Thus, in every region of Rindler spacetime, the Minkowski vacuum is replaced by pairs of entangled components. The direct-sum of all these components represents the entangled structure of the entire Rindler spacetime, which is a totally new picture in comparison \mbox{with \cite{Higuchi:2017gcd}} where unitarity is not maintained in each Rindler region like in our case. 

\section{Differences between Unruh Radiation in Standard QFT And DQFT}

\label{sec:DifferencesQFTs}

In standard quantum field theory, the derivation of Unruh radiation involves considering the behavior of a quantum field in the presence of the Rindler horizon. The field is quantized using standard techniques, and the resulting vacuum state for an inertial observer is equivalent to a thermal state for an accelerated observer. This thermal spectrum of particles, known as Unruh radiation, emerges due to the horizon's presence and the observer's acceleration.

In contrast, DQFT introduces a new framework for understanding quantum field theory in the context of horizons and superselection rules. In DQFT, the Hilbert space is decomposed into a direct-sum of sectorial Hilbert spaces associated with different regions of spacetimes separated by Horizons. 

The application of DQFT to the problem of Unruh radiation yields intriguing differences compared to standard QFT. Rather than considering a single vacuum state that evolves into a thermal state for accelerated observers, DQFT introduces the concept of pure states within the horizon for each Rindler observer. These pure states restore unitarity while simultaneously giving rise to a thermal spectrum of Unruh radiation.

Moreover, DQFT offers a new understanding of entanglement in Rindler spacetime. It suggests that Rindler observers can access complementary information beyond the horizon, challenging conventional notions of causality and information loss. The reinterpretation of the Reeh-Schlieder theorem within the framework of DQFT further underscores the novel insights this approach provides.

\section{Entanglement with Purity: An Essential Element for Unitarity Problem with Curved Spacetime} 

\label{sec:unitaritycurvedST}

This section draws analogies between DQFT in Rindler spacetime and DQFT in de Sitter and Schwarzchild spacetime earlier formulated in \cite{Kumar:2023ctp,Kumar:2023hbj}. We establish why DQFT is an essential element to understand a consistent QFT in curved spacetime that keeps the elements of entanglement but recovers pure states evolving into pure states even when the regions of spacetime are separated by horizons. 

Spacetime horizons are where gravity and QM mechanics act together. The Unruh effect is important in understanding quantum fields in curved spacetime. 
The best examples are de Sitter and Schwarzschild's black hole, where we can find a perfect analogy with Rindler spacetime. 

\subsection{DQFT in Schwarzchild Spacetime versus Rindler Spacetime}

\begin{figure}
    \includegraphics[width=0.6\linewidth]{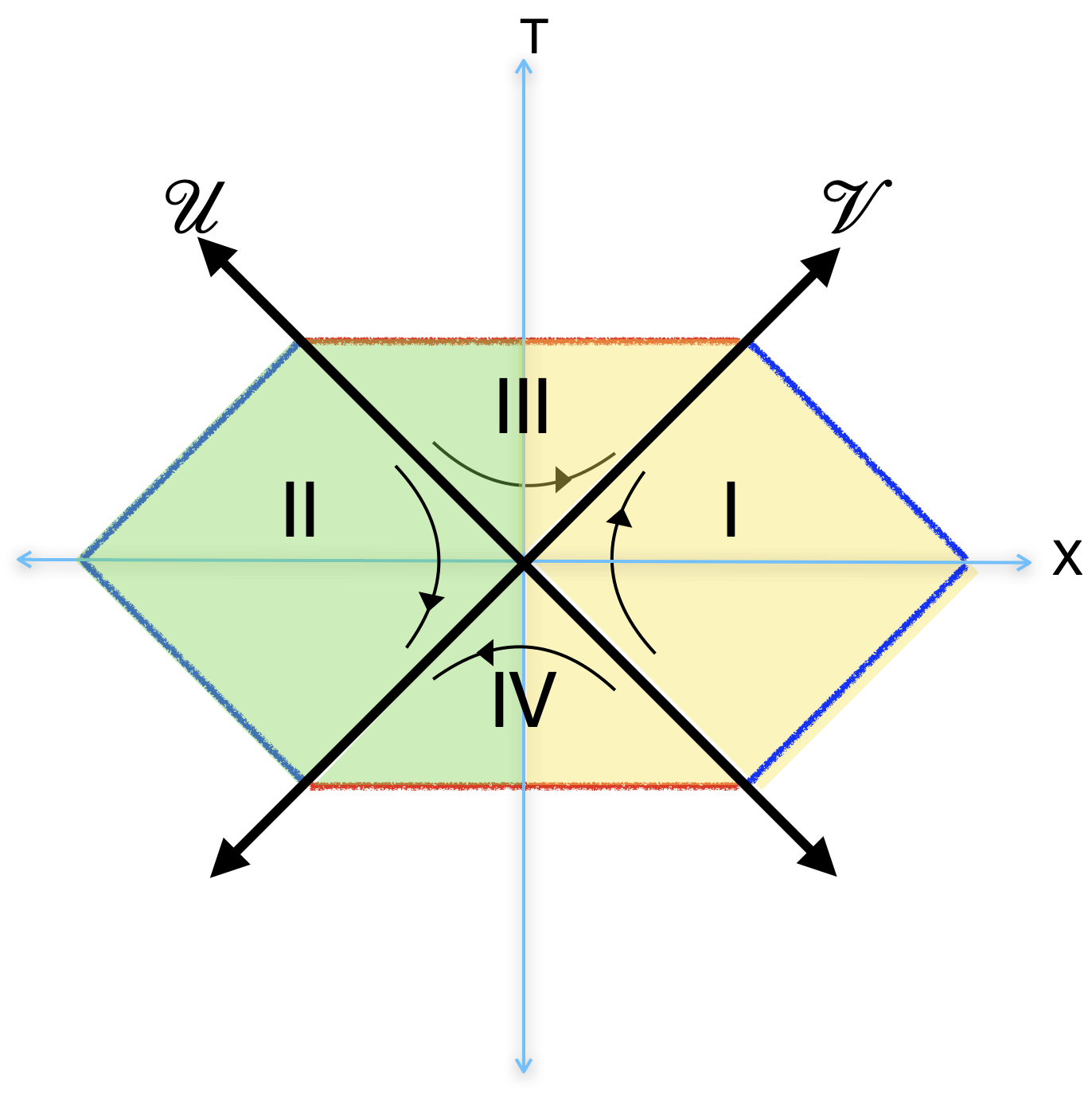}
    \caption{This is a conformal diagram of Schwarzschild black hole spacetime that represents a quantum field in DQFT formulation. 
    Here every point in yellow shaded region represents $\LF \theta,\,\varphi \RF$ whereas the green shaded region represents $\LF \pi-\theta,\,\pi+\varphi \RF$. Regions I and II represent the exterior static metric $r\gtrsim 2$ GM where, whereas Regions III and IV represent the interior Scharzchild metric. All these regions are to be interpreted quantum mechanically to depict a quantum field in superselection sectors. 
    The red lines in the diagram are identified, and they represent the singularity at $r=0$. 
    A quantum state in the entire region is a direct-sum of 4-components $ \vert \phi_{\rm total}\rangle = \frac{1}{\sqrt{2}} \LF \vert \phi_{I} \rangle \oplus \vert \phi_{II}\rangle \RF\oplus \frac{1}{\sqrt{2}} \LF \vert \phi_{III} \rangle\oplus \vert \phi_{IV}\rangle \RF$. }
    \label{fig:kssbh}
\end{figure}

Let us start with Schwarzschild black hole spacetime, which is usually written as
\begin{equation}
    ds^2 = -\LF 1-\frac{2GM}{r} \RF dt^2 + \frac{1}{\LF  1-\frac{2GM}{r}\RF }dr^2+ r^2d\Omega^2\,. 
\end{equation}
where $d\Omega^2 = d\theta^2+\sin^2\theta d\varphi^2 $ is the angular part of the metric. Schwarzchild spacetime is static (for $r\gtrsim 2GM$) and cosmological (for $r\lesssim 2$ GM) \cite{Doran:2006dq} (See also further discussion and extended list of references in \cite{Gaztanaga:2022fhp,gaztanaga2023a}). Thus, the horizon in Schwarzschild spacetime \mbox{is a boundary:}
\begin{enumerate}
    \item That separates cosmological and static spacetime 
    \item Where the Parity conjugate points on the Horizon 
    
    $\LF \theta,\,\varphi \RF\Big\vert_{r=2GM}$ and $\LF \pi-\theta,\,\pi+\varphi \RF\Big\vert_{r = 2}$ GM are space-like separated. 
\end{enumerate}
{Quantizing} 
a scalar field in Schwarzschild spacetime requires a new set of coordinates called Kruskal-Szekeres, defined by and depicted in Figure~\ref{fig:kssbh}.
\begin{equation}
    \begin{aligned}
       \Uc  &= -\kappa^{-1} e^{-\kappa \Tilde{u}}<0,\quad &&\Vc =\kappa^{-1} e^{\kappa \Tilde{v}}>0\quad &&(\rm Region\,I)\\
        \Uc &= \kappa^{-1} e^{-\kappa \Tilde{u}}>0,\quad &&\Vc = -\kappa^{-1} e^{\kappa \Tilde{v}}<0\quad &&(\rm Region\, II) \\
         \Uc &= \kappa^{-1} e^{-\kappa \Tilde{u}}>0,\quad &&\Vc= \kappa^{-1}e^{\kappa  \Tilde{v}}>0\quad &&(\rm Region\, III) \\
          \Uc &= -\kappa^{-1} e^{-\kappa \Tilde{u}}<0,\quad &&\Vc= -\kappa^{-1} e^{\kappa \Tilde{v}}<0\quad &&(\rm Region\, IV)
    \end{aligned}
    \label{UVKruskal}
\end{equation}
where $\kappa= \frac{1}{4\rm{GM}}$  is the surface gravity term. Here $\Tilde{u}= t-r_\ast$ and  $\Tilde{v}= t-r_\ast$ where $r_\ast$ being the so-called tortoise coordinate \cite{Griffiths:2009dfa}.
We can draw an analogy of these coordinates with the Rindler case \eqref{UVcoord} (ignoring the $\LF \theta,\,\varphi \RF$ part of the metric).

In Figure~\ref{fig:kssbh}, we depict the conformal diagram of Schwarzchild spacetime where components of a quantum state evolve in superselection sectors of the total Hilbert space. Further details and the purity of Hawking radiation can be found in \cite{Kumar:2023hbj}. The important message here is what we witness in the context of DQFT in Rindler spacetime is analogous to what we get in the context of DQFT in Schwarzschild spacetime (near $r\approx 2$ GM). According to this new hypothesis, quantum mechanically, Region III and Region IV of Figure~\ref{fig:kssbh} represent the entire interior of the Schwarzschild black hole, and there is no white hole. Similarly, Regions I and II represent the parity conjugate regions in the exterior of the Schwarzchild black hole, and there is no parallel universe. It is important to note that this is a completely quantum-mechanical description of spacetime. Once more, the concept of time in classical and quantum physics is not the same; ``time’’ is a parameter in quantum theory. Given this, we obtain a maximally entangled state that becomes a direct-sum of two components, which become pure states in the black hole's exterior and interior. Let us consider an entangled state, which is a pure state in the entire Hilbert space of (quantum) black hole spacetime (depicted in Figure~\ref{fig:kssbh}). According to the DQFT superselection rule, the horizon separates the pure state into two pure states of the superselection sectors, as we see below
\begin{equation}
    \vert \Tilde{\psi}_{12}\rangle =\sum_{m,n} \Tilde{c}_{mn} \vert \Tilde{\phi}_{m1}\rangle  \otimes \vert \Tilde{\phi}_{n2}\rangle = \frac{1}{\sqrt{2}}\sum_{m,n} \Tilde{c}_{mn} \LF \vert \Tilde{\phi}^{\rm ext}_{m1}\rangle  \otimes \Tilde{\phi}^{\rm ext}_{n2}\rangle \RF \oplus \LF \vert \Tilde{\phi}^{\rm int}_{m1}\rangle  \otimes \Tilde{\phi}^{\rm int}_{n2}\rangle \RF
\end{equation}
where $\Tilde{c}_{mn}\neq \Tilde{c}_n\Tilde{c}_m$, the superscripts ``ext’’ mean field component for $r\gtrsim 2GM$ whereas ``int’’ mean field component for $r\lesssim 2GM$ in the near horizon approximation.

DQFT construction in black hole spacetime preserves observer complementarity and unitarity similar to Rindler spacetime.  We drew here similarities between (quantum) black hole spacetime and (quantum) Rindler spacetime. Of course, differences exist because horizons in curved spacetime involve gravitational backreaction between interior and exterior states, unveiling the true nature of quantum gravity, which is worked out \mbox{in detail in \cite{Kumar:2023hbj}. }

\subsection{DQFT in Rindler Spacetime versus De Sitter Spacetime}

The importance of de Sitter (dS) spacetime is vast in cosmology; perfect examples are early Universe inflationary cosmology and late-time dark energy \cite{Starobinsky:2007hu,Starobinsky:1980te}.

We start with the dS metric in the flat Friedman-Lema\^itre-Robertson-Walker (FLRW) coordinates 
\begin{equation}
    ds^2 = -dt^2+e^{2Ht}d\textbf{x}^2 = \frac{1}{H^2\tau^2}\LF -d\tau^2+d\textbf{x}^2 \RF \,, 
    \label{dSmetric}
\end{equation}
which has the discrete symmetries 
\begin{equation}
    \LF \tau,\,\textbf{x}  \RF \longleftrightarrow  \LF -\tau,\,-\textbf{x}  \RF 
    \label{discsymds}
\end{equation}
similar to the $\Pc\Tc$ symmetry of Minkowski spacetime. It is often considered the metric \eqref{dSmetric} describes an expanding Universe for $\tau<0$, but the expanding Universe is characterized by the scale factor growth, not by the coordinate time. We can notice this in the \mbox{following equation} 
\begin{equation}
    a(t) = e^{Ht} \implies {\rm Expanding \,\, Universe} \implies \begin{cases}
        H>0 \quad t: -\infty \to \infty \\ 
        H<0 \quad t: \infty \to -\infty
    \end{cases}
\end{equation}
{Remember} 
that $H\to -H$ (the Hubble parameter $H=\frac{\dot{a}}{a}$ of de Sitter space) is the symmetry of de Sitter space (because curvature invariants are functions of $H^2$, for example, Ricci \mbox{scalar $R=12H^2$).} There are two arrows of time to characterize the expansion. In analogy, there are two arrows of time to define the positive energy state in standard QM as discussed in Section~\ref{sec:DQM}. Direct-sum quantum theory always combines the arrows of time with two parity conjugate regions to form superselection sectors. 

We can rewrite the metric \eqref{dSmetric} in the static coordinates $\LF t_s,\,r \RF$ as  \cite{Griffiths:2009dfa,nlab:de_sitter_spacetime}
 \begin{equation}
\begin{aligned}
    ds^2 & = -\LF 1-H^2r^2 \RF dt_s^2 + \frac{1}{\LF 1-H^2r^2 \RF}dr^2 + r^2d\Omega^2 \\ 
    &  = \frac{1}{H^2\LF 1-\Tilde{\Uc}\Tilde{\Vc} \RF^2} \LF -4d\Tilde{\Uc}d\Tilde{\Vc}+ \LF 1+\Tilde{\Uc}\Tilde{\Vc} \RF^2 d\Omega^2 \RF
    \end{aligned}
    \label{statdS}
\end{equation}
where $r= \big\vert \frac{1}{H}\big\vert \frac{1+\Tilde{\Uc}\Tilde{\Vc}}{1-\Tilde{\Uc}\Tilde{\Vc}}$ and $\frac{\Tilde{\Vc}}{\Tilde{\Uc}} = -e^{2Ht_s}$. The relation between the flat FLRW \eqref{dSmetric} and \eqref{statdS} is just a simple coordinate change \cite{Griffiths:2009dfa,nlab:de_sitter_spacetime}.
\begin{equation}
    r=r_fe^{Ht},\quad e^{-2Ht_s} = e^{-2Ht}-H^2r_f^2\,,
\end{equation}
where $r_f$ is the radial coordinate of flat FLRW dS \eqref{dSmetric} and the angular coordinates \mbox{$\Omega \equiv \LF \theta,\,\varphi \RF$} \linebreak remain the same in both metrics.

\begin{figure}
    \includegraphics[width=0.6\linewidth]{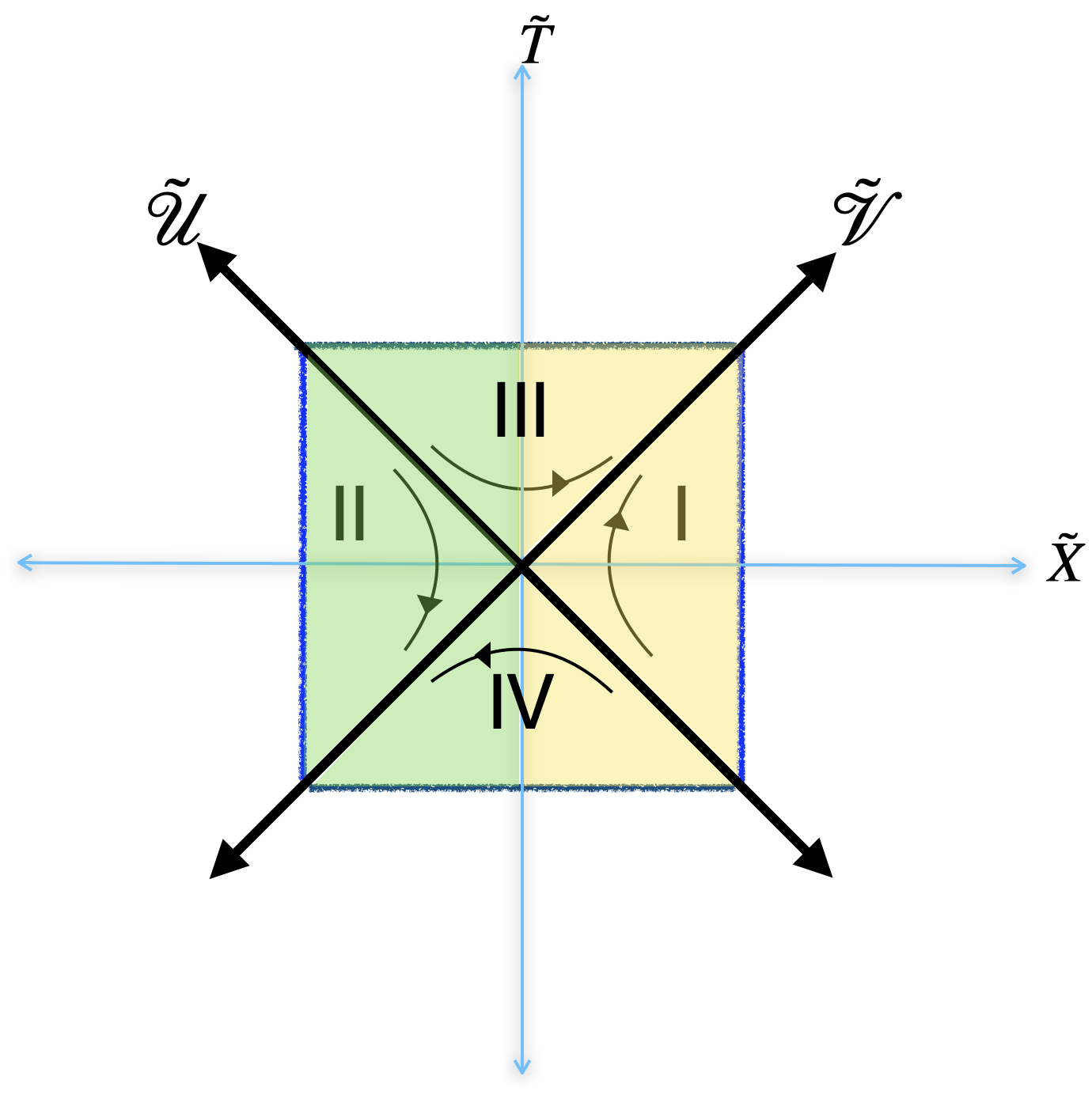}
    \caption{{Conformal} 
    diagram of de Sitter spacetime (in static coordinates) depicting the quantum states evolving in DQFT. Here also every point in yellow shaded region represents $\LF \theta,\,\varphi \RF$ whereas the green shaded region represents $\LF \pi-\theta,\,\pi+\varphi \RF$. The curvature of spacetime is constant everywhere as dS is maximally symmetric in nature. Analogous to Schwarzschild spacetime, a quantum state in the entire region is a direct-sum of 4-components $ \vert \phi_{\rm total}\rangle = \frac{1}{\sqrt{2}} \LF \vert \phi_{I} \rangle \oplus \vert \phi_{II}\rangle \RF\oplus \frac{1}{\sqrt{2}} \LF \vert \phi_{III} \rangle\oplus \vert \phi_{IV}\rangle \RF$ which leads to unitarity in dS spacetime.}
    \label{fig:ds}
\end{figure}

It is customary to consider in the literature \cite{nlab:de_sitter_spacetime} that \eqref{dSmetric} covers only half of the \mbox{dS space.} But taking into account the symmetry \eqref{discsymds}, dS in flat FLRW coordinates can cover the entire dS space where different regions are joined by discrete transformations. We can see this through four regions of dS spacetime (See Figure~\ref{fig:ds}) that the metric \eqref{statdS} represents and how they are related by discrete coordinate transformations. 

\begin{equation}
    \begin{aligned}
       \Uc  &= -e^{-H \bar{u}}<0,\quad &&\Vc= e^{H \bar{v}}>0\quad &&(\rm Region\,I)\\
        \Uc &= e^{-H \bar{u}}>0,\quad &&\Vc= -e^{H \bar{v}}<0\quad &&(\rm Region\, II) \\
         \Uc &= e^{-H \bar{u}}>0,\quad &&\Vc= e^{H \bar{v}}>0\quad &&(\rm Region\, III) \\
          \Uc &= -e^{-H \bar{u}}<0,\quad &&\Vc= -e^{H \bar{v}}<0\quad &&(\rm Region\, IV)
    \end{aligned}
    \label{UVKruskaldS}
\end{equation}
where $\bar u = t-\Tilde{r}_{\ast}$ and $\bar v = t+\Tilde{r}_{\ast}$ with $\Tilde{r}_\ast = \tanh^{-1}\LF H r_s \RF$. Figure~\ref{fig:ds} summarizes the DQFT description of dS spacetime, whose details can be found in \cite{Kumar:2023ctp}. We can again see an analogy of these coordinates \eqref{UVKruskaldS} with the Rindler case \eqref{UVcoord} (ignoring the $\LF \theta,\,\varphi \RF$ \mbox{part of the metric). }

We can compare all \eqref{UVcoord}, \eqref{UVKruskal}, and \eqref{UVKruskaldS}, and we can notice the similar structure despite the spacetimes being so different. We can also equivalently compare Figures~\ref{fig:RindlerST}--\ref{fig:ds} and deduce the universal features of the presence of horizons in both flat and curved spacetimes. 
This gives a universal picture of Rindler spacetime's importance in understanding quantum fields in curved spacetime. Unitary QFT in curved spacetime is the first step toward consistent quantum gravity. In this paper, we proposed a novel foundation for consistent quantum theory that can resolve the unitarity problem in the presence of spacetime horizons. Of course, there is still a long way to go for the full formulation of QFT in curved spacetime to address questions like $S-$matrix and scattering amplitudes, which are challenging subjects of investigation. All of this can be built successfully once the tenets of quantum theory, such as unitarity, are achieved. So, our attempt here is only to achieve unitarity. 
It is worth recalling that the DQFT construction of quantum fields in curved \mbox{spacetime \cite{Kumar:2023ctp}} has been successfully applied to inflationary quantum fluctuations that explained long-standing CMB anomalies with 650 times more effectively than the standard inflation \cite{Gaztanaga:2024vtr}. Thus, the theory we develop here has the needed observational evidence to move forward. 

\section{Conclusions and outlook}

\label{sec:conc}

In theoretical physics, the nature of information and its conservation within the universe remains a deeply intriguing question. Traditionally, it is believed that an observer's universe contains all possible information about events occurring within it, adhering to the principle of unitarity. Unitarity ensures that the evolution of a closed quantum system is governed by unitary operators, preserving the total information content over time. However, challenges to this notion arise in contexts involving horizons, such as black holes or cosmological event horizons, where the fate of information becomes ambiguous. For instance, the black hole information paradox raises questions about whether information falling into a black hole is irrevocably lost or can be recovered, suggesting a potential \mbox{breach of unitarity.}

In this study, we have investigated the implications of $\Pc\Tc$ symmetry in the context of Rindler spacetime and its effects on the perception of thermal radiation by Rindler observers. Our findings suggest a novel perspective on the nature of the thermal radiation detected by such observers.
Traditionally, the thermal radiation perceived by a Rindler observer is considered to be in a mixed state. This arises due to the Rindler horizon, which effectively partitions the spacetime into causally disconnected regions, leading to a loss of information about the state beyond the horizon. This approach leverages the inherent symmetry properties of spacetime, suggesting that the entanglement between states on either side of the Rindler horizon is not fundamentally lost but redistributed by \mbox{discrete transformations.}

In this paper, we have explored how direct-sum quantum field theory (DQFT) provides a robust framework for maintaining unitarity, even in the presence of such horizons. By decomposing the Hilbert space into a direct sum of sectorial Hilbert spaces, each associated with different superselection sectors, DQFT allows for a more nuanced understanding of information conservation. Interestingly, the DQFT mechanism of information reconstruction is not confined to Rindler spacetime alone. Similar principles can be applied to dS and Schwarzschild spacetimes. In dS and black hole spacetimes, where horizons introduce analogous challenges (similar to Rindler spacetime) in maintaining unitarity. The DQFT framework could allow information retrieval beyond the horizons by the rules of spiliting Hilbert or Fock spacetime into superselection sectors.

This study highlights a unifying theme in the treatment of horizons across different spacetimes. The superselection rules based on discrete transformations offer a powerful tool to preserve unitarity in the sectorial regions of spacetime separated by horizons. This perspective enriches our understanding of quantum field theory in uniformly accelerated frames. It has potential implications for broader contexts, such as black hole thermodynamics and the study of cosmological horizons. Further research in this direction may uncover deeper insights into the nature of spacetime symmetries and their role in the fundamental structure of quantum mechanics and gravity (quantum gravity).

Our findings suggest that within DQFT, each observer can access pure states within their respective horizons, restoring unitarity and ensuring that information is not lost. This approach reconciles the principles of unitarity with the observed phenomena, offering a potential resolution to long-standing paradoxes in theoretical physics.

By applying DQFT, we demonstrate that information does not leak beyond the confines of the observable universe. Still, rather, it is encoded in a way that respects the unitarity of quantum mechanics. 
 In conclusion, the direct-sum quantum field theory provides a promising mechanism for maintaining unitarity. It addresses the challenges posed by horizons and supports the notion that the universe, as observed by any observer, retains all information about its events. 



\vspace{6pt} 





\acknowledgments{KSK thanks the support of the Royal Society for the Newton International Fellowship. This research was funded by Funda\c{c}\~ao para a Ci\^encia e a Tecnologia grant number UIDB/MAT/00212/2020. We thank Enrique Gazta\~naga, Mathew Hull, Luca Buoninfante, Nava Gaddam, and Chandramouli Chowdhury for useful discussions. 
}


\appendix

\section{On the $\Pc\Tc$ symmetry of classical and quantum harmonic oscillator}  

\label{AppPT}

In this section, we discuss some basic details of our understanding of classical and quantum harmonic oscillators. From a physics point of view, quantum physics is superior to classical physics, but in reality and historically, we built quantum physics with a lot of input from classical physics. This section is organized in the space spirit. 

The Lagrangian of classical Harmonic oscillator is 
\begin{equation}
    \Lc_m = \frac{1}{2}m\LF \frac{dq_m}{dt} \RF^2 -\frac{m\tilde{\omega}^2 q_m^2}{2}=  \frac{p_m^2}{2m} -\frac{m\tilde{\omega}^2 q_m^2}{2}
\end{equation}
where $\LF p_m\,,q_m \RF$ are the momentum and position of the particle with mass $m$. The symmetry of the Lagrangian is 
\begin{equation}
    \LF p_m,\,q_m \RF \longleftrightarrow  \LF \bar{p}_m,\,\bar{q}_m \RF 
    \label{symPTc}
\end{equation}
where 
\begin{equation}
    \bar{p}_m = \frac{d\bar{q}_m}{d\bar{t}},\quad \bar{q}_m=-q_m,\quad \bar{t}=-t\,. 
\end{equation}
The symmetry \eqref{symPTc} indicates that one can consider space and time in terms of $\LF q_m,\,t \RF$ or $\LF \bar q_m,\,\bar t \RF$, the physics is invariant under which set of coordinates we choose. This is nothing but classical $\Pc\Tc$ symmetry. This should imply the construction of Schr\"{o}dinger equation should not also depend on what coordinates we choose, i.e., 
\begin{equation}
    i\frac{\pd \vert \Psi\rangle }{\pd t} = E\vert \Psi\rangle
    \label{sch1}
\end{equation}
should be equivalent to 
\begin{equation}
     i\frac{\pd \vert \Psi\rangle }{\pd \bar t} = -     i\frac{\pd \vert \Psi\rangle }{\pd t}  = E\vert \Psi\rangle
     \label{sch2}
\end{equation}
This should imply if \eqref{sch1} defines a positive energy state with respect to an arrow of time $t: -\infty \to \infty$, which should be the same as \eqref{sch2} defining a positive energy state with respect to an opposite arrow of time $t: \infty \to -\infty$. Thus, two definitions of a positive energy state depend on how we perceive an arrow of time. This happens because the standard Schr\"{o}dinger equation is a first-order differential equation in time. In contrast, the equation of motion of the classical harmonic oscillator is second-order in time. The direct-sum \eqref{disumSch} restores the $\Pc\Tc$ symmetry. By unifying $\Pc$ and $\Tc$, the definition of a positive energy state becomes disconnected from the arrow of time. A positive energy state now has two components at parity conjugate points that evolve with the opposite arrows of time.

\section{Bogoliubov transformations}

\label{appbogo}

In this section, we write down the general rules of Bogoliubov transformations \cite{Mukhanov:2007zz} when one expresses the field operators in two different basis $\LF u_k,\, u^\ast \RF$ and $\LF \bar{u}_p,\, \bar{u}_p^\ast \RF$ as linear combinations of each other with the respective creation and annihilation operators $\LF a_{\textbf{k}},\, a_{\textbf{k}}^\dagger \RF$ and $\LF b_{\textbf{p}},\, b_{\textbf{p}}^\dagger \RF$. 
\begin{equation}
\begin{aligned}
    u_k & = \int dp \LF \alpha_{kp}\bar{u}_p+ \beta_{kp}^\ast \bar{u}_p^\ast\RF  \\ 
    \bar{u}_p & = \int dp \LF \alpha_{kp}^\ast u_k - \beta_{kp}^\ast u_k^\ast \RF
    \end{aligned}\,,
    \label{modeBog}
\end{equation}
which follows from the Klein-Gordan inner product 
\begin{equation}
    \LF f,\,g \RF = i\int_\Sigma d\Sigma \LF \pd_{t_c} fg^\ast - \pd_{t_c}gf^\ast  \RF
\end{equation}
where $\Sigma$ is the spacelike Cauchy slice at time $t_c$. Here $\LF \alpha_{kp},\,\beta_{kp} \RF$ are the Bogoliubov coefficients which satisfy the following constraints dictated from canonical commutation relations 
\begin{equation}
    \begin{aligned}
        \int dp \LF \alpha^\ast_{kp}\alpha_{kp}- \beta_{kp}^\ast\beta_{kp} \RF & =1 \\ 
          \int dp \LF \alpha_{kp}\alpha_{kp}- \beta_{kp}\beta_{kp} \RF & = 0 \,.
    \end{aligned}
\end{equation}
From the Bogoliubov transformations \eqref{modeBog}, we can easily deduce the relation between the corresponding creation and annihilation operators as 
\begin{equation}
    \begin{pmatrix}
        \hat{b}_\textbf{p} \\ \hat{b}_\textbf{p}^\dagger  
    \end{pmatrix} = \int dk \begin{pmatrix}
        \alpha_{kp} & \beta_{kp} \\ 
        \beta_{kp}^\ast & \alpha_{kp}^\ast
    \end{pmatrix} \begin{pmatrix}
        \hat{a}_{\textbf{k}} \\ \hat{a}^\dagger_{\textbf{k}}
    \end{pmatrix}
    \label{CAbog}
\end{equation}
We can define here two vacuums with respect to the two sets of annihilation operators 
\begin{equation}
    a_\textbf{k}\vert 0_a\rangle =0 ,\quad b_\textbf{p}\vert 0_b\rangle =0\,.
\end{equation}
The Bogoliubov transformations \eqref{CAbog} imply 
\begin{equation}
b_\textbf{k}\vert 0_b\rangle =\int dk  \LF \alpha_{kp} a_\textbf{k} +\beta_{kp} a_\textbf{k}^\dagger \RF \vert 0_b\rangle  =0 
\end{equation}
This implies 
\begin{equation}
    a_\textbf{k} \vert 0\rangle_b = -\frac{\beta_{kp}}{\alpha_{kp}} a_\textbf{k}^\dagger\vert 0\rangle_b
\end{equation}
The solution to this equation  \cite{Jacobson:2003vx} can be obtained by turning it into a differential equation by substitution of $a_{\textbf{k}} = \frac{\pd}{\pd a^\dagger_{\textbf{k}}}$ that complies with the non-commutativity $\Big[ a_{\textbf{k}},\,a^\dagger_{\textbf{k}}\Big]=1$. Thus we obtain 
\begin{equation}
    \vert 0_b\rangle = \prod_{\textbf{p}}\frac{1}{\sqrt{\vert \alpha_{kp}}\vert} \exp\Bigg[-\LF\frac{\beta_{kp}}{2\alpha_{kp}^\ast}\RF a_\textbf{k}^\dagger a^\dagger_{-\textbf{k}}\Bigg] \vert 0_a\rangle
    \label{VACcom}
\end{equation}
The physical meaning of \eqref{VACcom} is that the vacuum $\vert 0_b\rangle$ appears to be excitation of vacuum $\vert 0_a\rangle$ with pairs of particles with opposite 3-momenta (due to isotropy of spacetime is assumed here \cite{Mukhanov:2007zz}).

We can compute the number density of b-particles in the vacuum $\vert 0_a\rangle$ as
\begin{equation}
    N_{b} = \langle 0_a\vert b^\dagger_\textbf{p} b_\textbf{p}\vert 0_a\rangle = \int dk \vert \beta_{kp}\vert^2\,. 
    \label{aveNb}
\end{equation}


\bibliography{References}

\begin{thebibliography}{37}%
\makeatletter
\providecommand \@ifxundefined [1]{%
 \@ifx{#1\undefined}
}%
\providecommand \@ifnum [1]{%
 \ifnum #1\expandafter \@firstoftwo
 \else \expandafter \@secondoftwo
 \fi
}%
\providecommand \@ifx [1]{%
 \ifx #1\expandafter \@firstoftwo
 \else \expandafter \@secondoftwo
 \fi
}%
\providecommand \natexlab [1]{#1}%
\providecommand \enquote  [1]{``#1''}%
\providecommand \bibnamefont  [1]{#1}%
\providecommand \bibfnamefont [1]{#1}%
\providecommand \citenamefont [1]{#1}%
\providecommand \href@noop [0]{\@secondoftwo}%
\providecommand \href [0]{\begingroup \@sanitize@url \@href}%
\providecommand \@href[1]{\@@startlink{#1}\@@href}%
\providecommand \@@href[1]{\endgroup#1\@@endlink}%
\providecommand \@sanitize@url [0]{\catcode `\\12\catcode `\$12\catcode `\&12\catcode `\#12\catcode `\^12\catcode `\_12\catcode `\%12\relax}%
\providecommand \@@startlink[1]{}%
\providecommand \@@endlink[0]{}%
\providecommand \url  [0]{\begingroup\@sanitize@url \@url }%
\providecommand \@url [1]{\endgroup\@href {#1}{\urlprefix }}%
\providecommand \urlprefix  [0]{URL }%
\providecommand \Eprint [0]{\href }%
\providecommand \doibase [0]{http://dx.doi.org/}%
\providecommand \selectlanguage [0]{\@gobble}%
\providecommand \bibinfo  [0]{\@secondoftwo}%
\providecommand \bibfield  [0]{\@secondoftwo}%
\providecommand \translation [1]{[#1]}%
\providecommand \BibitemOpen [0]{}%
\providecommand \bibitemStop [0]{}%
\providecommand \bibitemNoStop [0]{.\EOS\space}%
\providecommand \EOS [0]{\spacefactor3000\relax}%
\providecommand \BibitemShut  [1]{\csname bibitem#1\endcsname}%
\let\auto@bib@innerbib\@empty
\bibitem [{\citenamefont {Hawking}(1975)}]{Hawking:1974sw}%
  \BibitemOpen
  \bibfield  {author} {\bibinfo {author} {\bibfnamefont {S.~W.}\ \bibnamefont {Hawking}},\ }\bibfield  {booktitle} {\emph {\bibinfo {booktitle} {{Euclidean quantum gravity}}},\ }\href {\doibase 10.1007/BF02345020} {\bibfield  {journal} {\bibinfo  {journal} {Commun. Math. Phys.}\ }\textbf {\bibinfo {volume} {43}},\ \bibinfo {pages} {199} (\bibinfo {year} {1975})},\ \bibinfo {note} {[,167(1975)]}\BibitemShut {NoStop}%
\bibitem [{\citenamefont {Kumar}\ and\ \citenamefont {Marto}(2024)}]{Kumar:2023hbj}%
  \BibitemOpen
  \bibfield  {author} {\bibinfo {author} {\bibfnamefont {K.~S.}\ \bibnamefont {Kumar}}\ and\ \bibinfo {author} {\bibfnamefont {J.}~\bibnamefont {Marto}},\ }\href {\doibase 10.1093/ptep/ptae176} {\bibfield  {journal} {\bibinfo  {journal} {PTEP}\ }\textbf {\bibinfo {volume} {ptae}},\ \bibinfo {pages} {176} (\bibinfo {year} {2024})},\ \Eprint {http://arxiv.org/abs/2307.10345} {arXiv:2307.10345 [hep-th]} \BibitemShut {NoStop}%
\bibitem [{\citenamefont {Unruh}(1976)}]{Unruh1976@article}%
  \BibitemOpen
  \bibfield  {author} {\bibinfo {author} {\bibfnamefont {W.~G.}\ \bibnamefont {Unruh}},\ }\href {\doibase 10.1103/PhysRevD.14.870} {\bibfield  {journal} {\bibinfo  {journal} {Phys. Rev. D}\ }\textbf {\bibinfo {volume} {14}},\ \bibinfo {pages} {870} (\bibinfo {year} {1976})}\BibitemShut {NoStop}%
\bibitem [{\citenamefont {Higuchi}\ \emph {et~al.}(2017)\citenamefont {Higuchi}, \citenamefont {Iso}, \citenamefont {Ueda},\ and\ \citenamefont {Yamamoto}}]{Higuchi:2017gcd}%
  \BibitemOpen
  \bibfield  {author} {\bibinfo {author} {\bibfnamefont {A.}~\bibnamefont {Higuchi}}, \bibinfo {author} {\bibfnamefont {S.}~\bibnamefont {Iso}}, \bibinfo {author} {\bibfnamefont {K.}~\bibnamefont {Ueda}}, \ and\ \bibinfo {author} {\bibfnamefont {K.}~\bibnamefont {Yamamoto}},\ }\href {\doibase 10.1103/PhysRevD.96.083531} {\bibfield  {journal} {\bibinfo  {journal} {Phys. Rev. D}\ }\textbf {\bibinfo {volume} {96}},\ \bibinfo {pages} {083531} (\bibinfo {year} {2017})},\ \Eprint {http://arxiv.org/abs/1709.05757} {arXiv:1709.05757 [hep-th]} \BibitemShut {NoStop}%
\bibitem [{\citenamefont {Crispino}\ \emph {et~al.}(2008)\citenamefont {Crispino}, \citenamefont {Higuchi},\ and\ \citenamefont {Matsas}}]{Crispino:2007eb}%
  \BibitemOpen
  \bibfield  {author} {\bibinfo {author} {\bibfnamefont {L.~C.~B.}\ \bibnamefont {Crispino}}, \bibinfo {author} {\bibfnamefont {A.}~\bibnamefont {Higuchi}}, \ and\ \bibinfo {author} {\bibfnamefont {G.~E.~A.}\ \bibnamefont {Matsas}},\ }\href {\doibase 10.1103/RevModPhys.80.787} {\bibfield  {journal} {\bibinfo  {journal} {Rev. Mod. Phys.}\ }\textbf {\bibinfo {volume} {80}},\ \bibinfo {pages} {787} (\bibinfo {year} {2008})},\ \Eprint {http://arxiv.org/abs/0710.5373} {arXiv:0710.5373 [gr-qc]} \BibitemShut {NoStop}%
\bibitem [{\citenamefont {Mukhanov}\ and\ \citenamefont {Winitzki}(2007)}]{Mukhanov:2007zz}%
  \BibitemOpen
  \bibfield  {author} {\bibinfo {author} {\bibfnamefont {V.}~\bibnamefont {Mukhanov}}\ and\ \bibinfo {author} {\bibfnamefont {S.}~\bibnamefont {Winitzki}},\ }\href@noop {} {\emph {\bibinfo {title} {{Introduction to quantum effects in gravity}}}}\ (\bibinfo  {publisher} {Cambridge University Press},\ \bibinfo {year} {2007})\BibitemShut {NoStop}%
\bibitem [{\citenamefont {Raju}(2022)}]{Raju:2020smc}%
  \BibitemOpen
  \bibfield  {author} {\bibinfo {author} {\bibfnamefont {S.}~\bibnamefont {Raju}},\ }\href {\doibase 10.1016/j.physrep.2021.10.001} {\bibfield  {journal} {\bibinfo  {journal} {Phys. Rept.}\ }\textbf {\bibinfo {volume} {943}},\ \bibinfo {pages} {1} (\bibinfo {year} {2022})},\ \Eprint {http://arxiv.org/abs/2012.05770} {arXiv:2012.05770 [hep-th]} \BibitemShut {NoStop}%
\bibitem [{\citenamefont {Almheiri}\ \emph {et~al.}(2020)\citenamefont {Almheiri}, \citenamefont {Mahajan}, \citenamefont {Maldacena},\ and\ \citenamefont {Zhao}}]{Almheiri_2020}%
  \BibitemOpen
  \bibfield  {author} {\bibinfo {author} {\bibfnamefont {A.}~\bibnamefont {Almheiri}}, \bibinfo {author} {\bibfnamefont {R.}~\bibnamefont {Mahajan}}, \bibinfo {author} {\bibfnamefont {J.}~\bibnamefont {Maldacena}}, \ and\ \bibinfo {author} {\bibfnamefont {Y.}~\bibnamefont {Zhao}},\ }\href {\doibase 10.1007/jhep03(2020)149} {\bibfield  {journal} {\bibinfo  {journal} {Journal of High Energy Physics}\ }\textbf {\bibinfo {volume} {2020}} (\bibinfo {year} {2020}),\ 10.1007/jhep03(2020)149}\BibitemShut {NoStop}%
\bibitem [{\citenamefont {Hawking}\ \emph {et~al.}(2016)\citenamefont {Hawking}, \citenamefont {Perry},\ and\ \citenamefont {Strominger}}]{Hawking:2016msc}%
  \BibitemOpen
  \bibfield  {author} {\bibinfo {author} {\bibfnamefont {S.~W.}\ \bibnamefont {Hawking}}, \bibinfo {author} {\bibfnamefont {M.~J.}\ \bibnamefont {Perry}}, \ and\ \bibinfo {author} {\bibfnamefont {A.}~\bibnamefont {Strominger}},\ }\href {\doibase 10.1103/PhysRevLett.116.231301} {\bibfield  {journal} {\bibinfo  {journal} {Phys. Rev. Lett.}\ }\textbf {\bibinfo {volume} {116}},\ \bibinfo {pages} {231301} (\bibinfo {year} {2016})},\ \Eprint {http://arxiv.org/abs/1601.00921} {arXiv:1601.00921 [hep-th]} \BibitemShut {NoStop}%
\bibitem [{\citenamefont {Sanchez}\ and\ \citenamefont {Whiting}(1987)}]{SANCHEZ1987605}%
  \BibitemOpen
  \bibfield  {author} {\bibinfo {author} {\bibfnamefont {N.}~\bibnamefont {Sanchez}}\ and\ \bibinfo {author} {\bibfnamefont {B.}~\bibnamefont {Whiting}},\ }\href {\doibase https://doi.org/10.1016/0550-3213(87)90289-6} {\bibfield  {journal} {\bibinfo  {journal} {Nuclear Physics B}\ }\textbf {\bibinfo {volume} {283}},\ \bibinfo {pages} {605} (\bibinfo {year} {1987})}\BibitemShut {NoStop}%
\bibitem [{\citenamefont {Sánchez}(1987)}]{SANCHEZ19871111}%
  \BibitemOpen
  \bibfield  {author} {\bibinfo {author} {\bibfnamefont {N.}~\bibnamefont {Sánchez}},\ }\href {\doibase https://doi.org/10.1016/0550-3213(87)90625-0} {\bibfield  {journal} {\bibinfo  {journal} {Nuclear Physics B}\ }\textbf {\bibinfo {volume} {294}},\ \bibinfo {pages} {1111} (\bibinfo {year} {1987})}\BibitemShut {NoStop}%
\bibitem [{\citenamefont {Schr{\"o}dinger}(1956)}]{Schrodinger1956}%
  \BibitemOpen
  \bibfield  {author} {\bibinfo {author} {\bibfnamefont {E.}~\bibnamefont {Schr{\"o}dinger}},\ }\href {https://books.google.co.uk/books?id=UQTsuQEACAAJ} {\emph {\bibinfo {title} {Expanding Universe}}}\ (\bibinfo  {publisher} {Cambridge University Press},\ \bibinfo {year} {1956})\BibitemShut {NoStop}%
\bibitem [{\citenamefont {'t~Hooft}(2017)}]{tHooft:2016rrl}%
  \BibitemOpen
  \bibfield  {author} {\bibinfo {author} {\bibfnamefont {G.}~\bibnamefont {'t~Hooft}},\ }\href {\doibase 10.1007/s10701-017-0122-3} {\bibfield  {journal} {\bibinfo  {journal} {Found. Phys.}\ }\textbf {\bibinfo {volume} {47}},\ \bibinfo {pages} {1503} (\bibinfo {year} {2017})},\ \Eprint {http://arxiv.org/abs/1612.08640} {arXiv:1612.08640 [gr-qc]} \BibitemShut {NoStop}%
\bibitem [{\citenamefont {'t~Hooft}(1993)}]{tHooft:1993dmi}%
  \BibitemOpen
  \bibfield  {author} {\bibinfo {author} {\bibfnamefont {G.}~\bibnamefont {'t~Hooft}},\ }\bibfield  {booktitle} {\emph {\bibinfo {booktitle} {{Conference on Highlights of Particle and Condensed Matter Physics (SALAMFEST) Trieste, Italy, March 8-12, 1993}}},\ }\href@noop {} {\bibfield  {journal} {\bibinfo  {journal} {Conf. Proc.}\ }\textbf {\bibinfo {volume} {C930308}},\ \bibinfo {pages} {284} (\bibinfo {year} {1993})},\ \Eprint {http://arxiv.org/abs/gr-qc/9310026} {arXiv:gr-qc/9310026 [gr-qc]} \BibitemShut {NoStop}%
\bibitem [{\citenamefont {'t~Hooft}(2018)}]{tHooft:2018waj}%
  \BibitemOpen
  \bibfield  {author} {\bibinfo {author} {\bibfnamefont {G.}~\bibnamefont {'t~Hooft}},\ }\href {\doibase 10.1007/s10701-017-0133-0} {\bibfield  {journal} {\bibinfo  {journal} {Found. Phys.}\ }\textbf {\bibinfo {volume} {48}},\ \bibinfo {pages} {1134} (\bibinfo {year} {2018})}\BibitemShut {NoStop}%
\bibitem [{\citenamefont {{nLab authors}}(2023)}]{nlab:superselection_theory}%
  \BibitemOpen
  \bibfield  {author} {\bibinfo {author} {\bibnamefont {{nLab authors}}},\ }\href@noop {} {\enquote {\bibinfo {title} {superselection theory},}\ }\bibinfo {howpublished} {\url{https://ncatlab.org/nlab/show/superselection+theory}} (\bibinfo {year} {2023}),\ \bibinfo {note} {\href{https://ncatlab.org/nlab/revision/superselection+theory/10}{Revision 10}}\BibitemShut {NoStop}%
\bibitem [{\citenamefont {Kumar}\ and\ \citenamefont {Marto}(2023)}]{Kumar:2023ctp}%
  \BibitemOpen
  \bibfield  {author} {\bibinfo {author} {\bibfnamefont {K.~S.}\ \bibnamefont {Kumar}}\ and\ \bibinfo {author} {\bibfnamefont {J.}~\bibnamefont {Marto}},\ }\href@noop {} {\  (\bibinfo {year} {2023})},\ \Eprint {http://arxiv.org/abs/2305.06046} {arXiv:2305.06046 [hep-th]} \BibitemShut {NoStop}%
\bibitem [{\citenamefont {Gazta\~naga}\ and\ \citenamefont {Kumar}(2024)}]{Gaztanaga:2024vtr}%
  \BibitemOpen
  \bibfield  {author} {\bibinfo {author} {\bibfnamefont {E.}~\bibnamefont {Gazta\~naga}}\ and\ \bibinfo {author} {\bibfnamefont {K.~S.}\ \bibnamefont {Kumar}},\ }\href {\doibase 10.1088/1475-7516/2024/06/001} {\bibfield  {journal} {\bibinfo  {journal} {JCAP}\ }\textbf {\bibinfo {volume} {06}},\ \bibinfo {pages} {001} (\bibinfo {year} {2024})},\ \Eprint {http://arxiv.org/abs/2401.08288} {arXiv:2401.08288 [astro-ph.CO]} \BibitemShut {NoStop}%
\bibitem [{\citenamefont {Simon}\ \emph {et~al.}(2008)\citenamefont {Simon}, \citenamefont {Mukunda}, \citenamefont {Chaturvedi},\ and\ \citenamefont {Srinivasan}}]{Simon_2008}%
  \BibitemOpen
  \bibfield  {author} {\bibinfo {author} {\bibfnamefont {R.}~\bibnamefont {Simon}}, \bibinfo {author} {\bibfnamefont {N.}~\bibnamefont {Mukunda}}, \bibinfo {author} {\bibfnamefont {S.}~\bibnamefont {Chaturvedi}}, \ and\ \bibinfo {author} {\bibfnamefont {V.}~\bibnamefont {Srinivasan}},\ }\href {\doibase 10.1016/j.physleta.2008.09.052} {\bibfield  {journal} {\bibinfo  {journal} {Physics Letters A}\ }\textbf {\bibinfo {volume} {372}},\ \bibinfo {pages} {6847–6852} (\bibinfo {year} {2008})}\BibitemShut {NoStop}%
\bibitem [{\citenamefont {Roberts}(2022)}]{Roberts:2022xcj}%
  \BibitemOpen
  \bibfield  {author} {\bibinfo {author} {\bibfnamefont {B.~W.}\ \bibnamefont {Roberts}},\ }\href {\doibase 10.1017/9781009122139} {\  (\bibinfo {year} {2022}),\ 10.1017/9781009122139},\ \Eprint {http://arxiv.org/abs/2212.03489} {arXiv:2212.03489 [physics.hist-ph]} \BibitemShut {NoStop}%
\bibitem [{\citenamefont {Srednicki}(2007)}]{Srednicki:2007qs}%
  \BibitemOpen
  \bibfield  {author} {\bibinfo {author} {\bibfnamefont {M.}~\bibnamefont {Srednicki}},\ }\href@noop {} {\emph {\bibinfo {title} {{Quantum field theory}}}}\ (\bibinfo  {publisher} {Cambridge University Press},\ \bibinfo {year} {2007})\BibitemShut {NoStop}%
\bibitem [{\citenamefont {Coleman}(2018)}]{Coleman:2018mew}%
  \BibitemOpen
  \bibfield  {author} {\bibinfo {author} {\bibfnamefont {S.}~\bibnamefont {Coleman}},\ }\href {\doibase 10.1142/9371} {\emph {\bibinfo {title} {{Lectures of Sidney Coleman on Quantum Field Theory}}}},\ edited by\ \bibinfo {editor} {\bibfnamefont {B.~G.-g.}\ \bibnamefont {Chen}}, \bibinfo {editor} {\bibfnamefont {D.}~\bibnamefont {Derbes}}, \bibinfo {editor} {\bibfnamefont {D.}~\bibnamefont {Griffiths}}, \bibinfo {editor} {\bibfnamefont {B.}~\bibnamefont {Hill}}, \bibinfo {editor} {\bibfnamefont {R.}~\bibnamefont {Sohn}}, \ and\ \bibinfo {editor} {\bibfnamefont {Y.-S.}\ \bibnamefont {Ting}}\ (\bibinfo  {publisher} {WSP},\ \bibinfo {address} {Hackensack},\ \bibinfo {year} {2018})\BibitemShut {NoStop}%
\bibitem [{\citenamefont {Aharonov}\ and\ \citenamefont {Vaidman}(2002)}]{Aharonov2002}%
  \BibitemOpen
  \bibfield  {author} {\bibinfo {author} {\bibfnamefont {Y.}~\bibnamefont {Aharonov}}\ and\ \bibinfo {author} {\bibfnamefont {L.}~\bibnamefont {Vaidman}},\ }\enquote {\bibinfo {title} {The two-state vector formalism of quantum mechanics},}\ in\ \href {\doibase 10.1007/3-540-45846-8_13} {\emph {\bibinfo {booktitle} {Time in Quantum Mechanics}}},\ \bibinfo {editor} {edited by\ \bibinfo {editor} {\bibfnamefont {J.~G.}\ \bibnamefont {Muga}}, \bibinfo {editor} {\bibfnamefont {R.~S.}\ \bibnamefont {Mayato}}, \ and\ \bibinfo {editor} {\bibfnamefont {I.~L.}\ \bibnamefont {Egusquiza}}}\ (\bibinfo  {publisher} {Springer Berlin Heidelberg},\ \bibinfo {address} {Berlin, Heidelberg},\ \bibinfo {year} {2002})\ pp.\ \bibinfo {pages} {369--412}\BibitemShut {NoStop}%
\bibitem [{\citenamefont {Conway}(2010)}]{Conway}%
  \BibitemOpen
  \bibfield  {author} {\bibinfo {author} {\bibfnamefont {J.~B.}\ \bibnamefont {Conway}},\ }\href@noop {} {\emph {\bibinfo {title} {A course in functional analysis}}},\ \bibinfo {edition} {2nd}\ ed.,\ Graduate texts in mathematics ; 96\ (\bibinfo  {publisher} {Springer Science+Business Media},\ \bibinfo {address} {New York},\ \bibinfo {year} {2010})\BibitemShut {NoStop}%
\bibitem [{\citenamefont {Rovelli}(2004)}]{rovelli_2004}%
  \BibitemOpen
  \bibfield  {author} {\bibinfo {author} {\bibfnamefont {C.}~\bibnamefont {Rovelli}},\ }\href {\doibase 10.1017/CBO9780511755804} {\emph {\bibinfo {title} {Quantum Gravity}}},\ Cambridge Monographs on Mathematical Physics\ (\bibinfo  {publisher} {Cambridge University Press},\ \bibinfo {year} {2004})\BibitemShut {NoStop}%
\bibitem [{\citenamefont {Reeh}\ and\ \citenamefont {Schlieder}(1961)}]{Reeh:1961ujh}%
  \BibitemOpen
  \bibfield  {author} {\bibinfo {author} {\bibfnamefont {H.}~\bibnamefont {Reeh}}\ and\ \bibinfo {author} {\bibfnamefont {S.}~\bibnamefont {Schlieder}},\ }\href {\doibase 10.1007/BF02787889} {\bibfield  {journal} {\bibinfo  {journal} {Nuovo Cim.}\ }\textbf {\bibinfo {volume} {22}},\ \bibinfo {pages} {1051} (\bibinfo {year} {1961})}\BibitemShut {NoStop}%
\bibitem [{\citenamefont {Witten}(2018)}]{Witten:2018zxz}%
  \BibitemOpen
  \bibfield  {author} {\bibinfo {author} {\bibfnamefont {E.}~\bibnamefont {Witten}},\ }\href {\doibase 10.1103/RevModPhys.90.045003} {\bibfield  {journal} {\bibinfo  {journal} {Rev. Mod. Phys.}\ }\textbf {\bibinfo {volume} {90}},\ \bibinfo {pages} {045003} (\bibinfo {year} {2018})},\ \Eprint {http://arxiv.org/abs/1803.04993} {arXiv:1803.04993 [hep-th]} \BibitemShut {NoStop}%
\bibitem [{\citenamefont {Agullo}\ \emph {et~al.}(2023)\citenamefont {Agullo}, \citenamefont {Bonga}, \citenamefont {Ribes-Metidieri}, \citenamefont {Kranas},\ and\ \citenamefont {Nadal-Gisbert}}]{Agullo:2023fnp}%
  \BibitemOpen
  \bibfield  {author} {\bibinfo {author} {\bibfnamefont {I.}~\bibnamefont {Agullo}}, \bibinfo {author} {\bibfnamefont {B.}~\bibnamefont {Bonga}}, \bibinfo {author} {\bibfnamefont {P.}~\bibnamefont {Ribes-Metidieri}}, \bibinfo {author} {\bibfnamefont {D.}~\bibnamefont {Kranas}}, \ and\ \bibinfo {author} {\bibfnamefont {S.}~\bibnamefont {Nadal-Gisbert}},\ }\href {\doibase 10.1103/PhysRevD.108.085005} {\bibfield  {journal} {\bibinfo  {journal} {Phys. Rev. D}\ }\textbf {\bibinfo {volume} {108}},\ \bibinfo {pages} {085005} (\bibinfo {year} {2023})},\ \Eprint {http://arxiv.org/abs/2302.13742} {arXiv:2302.13742 [quant-ph]} \BibitemShut {NoStop}%
\bibitem [{\citenamefont {Haag}(1992)}]{Haag:1992hx}%
  \BibitemOpen
  \bibfield  {author} {\bibinfo {author} {\bibfnamefont {R.}~\bibnamefont {Haag}},\ }\href@noop {} {\emph {\bibinfo {title} {{Local quantum physics: Fields, particles, algebras}}}}\ (\bibinfo {year} {1992})\BibitemShut {NoStop}%
\bibitem [{\citenamefont {Doran}\ \emph {et~al.}(2008)\citenamefont {Doran}, \citenamefont {Lobo},\ and\ \citenamefont {Crawford}}]{Doran:2006dq}%
  \BibitemOpen
  \bibfield  {author} {\bibinfo {author} {\bibfnamefont {R.}~\bibnamefont {Doran}}, \bibinfo {author} {\bibfnamefont {F.~S.~N.}\ \bibnamefont {Lobo}}, \ and\ \bibinfo {author} {\bibfnamefont {P.}~\bibnamefont {Crawford}},\ }\href {\doibase 10.1007/s10701-007-9197-6} {\bibfield  {journal} {\bibinfo  {journal} {Found. Phys.}\ }\textbf {\bibinfo {volume} {38}},\ \bibinfo {pages} {160} (\bibinfo {year} {2008})},\ \Eprint {http://arxiv.org/abs/gr-qc/0609042} {arXiv:gr-qc/0609042} \BibitemShut {NoStop}%
\bibitem [{\citenamefont {Gaztanaga}(2022)}]{Gaztanaga:2022fhp}%
  \BibitemOpen
  \bibfield  {author} {\bibinfo {author} {\bibfnamefont {E.}~\bibnamefont {Gaztanaga}},\ }\href {\doibase 10.3390/sym14091849} {\bibfield  {journal} {\bibinfo  {journal} {Symmetry}\ }\textbf {\bibinfo {volume} {14}},\ \bibinfo {pages} {1849} (\bibinfo {year} {2022})}\BibitemShut {NoStop}%
\bibitem [{\citenamefont {{Gazta{\~n}aga}}(2023)}]{gaztanaga2023a}%
  \BibitemOpen
  \bibfield  {author} {\bibinfo {author} {\bibfnamefont {E.}~\bibnamefont {{Gazta{\~n}aga}}},\ }\href {\doibase 10.1093/mnrasl/slad015} {\bibfield  {journal} {\bibinfo  {journal} {MNRAS}\ }\textbf {\bibinfo {volume} {521}},\ \bibinfo {pages} {L59} (\bibinfo {year} {2023})}\BibitemShut {NoStop}%
\bibitem [{\citenamefont {Griffiths}\ and\ \citenamefont {Podolsky}(2009)}]{Griffiths:2009dfa}%
  \BibitemOpen
  \bibfield  {author} {\bibinfo {author} {\bibfnamefont {J.~B.}\ \bibnamefont {Griffiths}}\ and\ \bibinfo {author} {\bibfnamefont {J.}~\bibnamefont {Podolsky}},\ }\href {\doibase 10.1017/CBO9780511635397} {\emph {\bibinfo {title} {{Exact Space-Times in Einstein's General Relativity}}}},\ Cambridge Monographs on Mathematical Physics\ (\bibinfo  {publisher} {Cambridge University Press},\ \bibinfo {address} {Cambridge},\ \bibinfo {year} {2009})\BibitemShut {NoStop}%
\bibitem [{\citenamefont {Starobinsky}(2007)}]{Starobinsky:2007hu}%
  \BibitemOpen
  \bibfield  {author} {\bibinfo {author} {\bibfnamefont {A.~A.}\ \bibnamefont {Starobinsky}},\ }\href {\doibase 10.1134/S0021364007150027} {\bibfield  {journal} {\bibinfo  {journal} {JETP Lett.}\ }\textbf {\bibinfo {volume} {86}},\ \bibinfo {pages} {157} (\bibinfo {year} {2007})},\ \Eprint {http://arxiv.org/abs/0706.2041} {arXiv:0706.2041 [astro-ph]} \BibitemShut {NoStop}%
\bibitem [{\citenamefont {Starobinsky}(1980)}]{Starobinsky:1980te}%
  \BibitemOpen
  \bibfield  {author} {\bibinfo {author} {\bibfnamefont {A.~A.}\ \bibnamefont {Starobinsky}},\ }\href {\doibase 10.1016/0370-2693(80)90670-X} {\bibfield  {journal} {\bibinfo  {journal} {Phys. Lett. B}\ }\textbf {\bibinfo {volume} {91}},\ \bibinfo {pages} {99} (\bibinfo {year} {1980})}\BibitemShut {NoStop}%
\bibitem [{\citenamefont {{nLab authors}}(2024)}]{nlab:de_sitter_spacetime}%
  \BibitemOpen
  \bibfield  {author} {\bibinfo {author} {\bibnamefont {{nLab authors}}},\ }\href@noop {} {\enquote {\bibinfo {title} {de {{S}}itter spacetime},}\ }\bibinfo {howpublished} {\url{https://ncatlab.org/nlab/show/de+Sitter+spacetime}} (\bibinfo {year} {2024}),\ \bibinfo {note} {\href{https://ncatlab.org/nlab/revision/de+Sitter+spacetime/43}{Revision 43}}\BibitemShut {NoStop}%
\bibitem [{\citenamefont {Jacobson}(2003)}]{Jacobson:2003vx}%
  \BibitemOpen
  \bibfield  {author} {\bibinfo {author} {\bibfnamefont {T.}~\bibnamefont {Jacobson}},\ }in\ \href {\doibase 10.1007/0-387-24992-3_2} {\emph {\bibinfo {booktitle} {{School on Quantum Gravity}}}}\ (\bibinfo {year} {2003})\ pp.\ \bibinfo {pages} {39--89},\ \Eprint {http://arxiv.org/abs/gr-qc/0308048} {arXiv:gr-qc/0308048} \BibitemShut {NoStop}%
\end{thebibliography}%

\end{document}